\documentclass[chicago]{emulateapj}
\usepackage{amsmath}
\usepackage{multirow}
\usepackage[dvips]{color}

\slugcomment{submit v2}
\shortauthors{}

\begin{document}

\title{Time Monitoring of Radio Jets and Magnetospheres in the Nearby Young Stellar Cluster R Coronae Australis}

\author{Hauyu Baobab Liu\altaffilmark{1}} \author{Roberto Galv\'{a}n-Madrid\altaffilmark{2}} \author{Jan Forbrich\altaffilmark{3}} \author{Luis F. Rodr\'{i}guez\altaffilmark{4}} \author{Michihiro Takami\altaffilmark{1}} \author{Gr\'{a}inne Costigan\altaffilmark{2}} \author{Carlo Felice Manara\altaffilmark{2}}  \author{Chi-Hung Yan\altaffilmark{1,7}} \author{Jennifer Karr\altaffilmark{1}} \author{Mei-Yin Chou\altaffilmark{1}} \author{Paul T.-P. Ho\altaffilmark{1,8}} \author{Qizhou Zhang\altaffilmark{8}}

\affil{$^{1}$Academia Sinica Institute of Astronomy and Astrophysics, P.O. Box 23-141, Taipei, 106 Taiwan} \email{hyliu@asiaa.sinica.edu.tw}
\affil{$^{2}$European Southern Observatory, Karl-Schwarzschild-Str. 2, 85748, Garching, Germany}
\affil{$^{3}$University of Vienna, Department of Astrophysics, T\"{u}rkenschanzstra$\ss$e 17, 1180, Vienna, Austria}
\affil{$^{4}$Centro de Radioastronom\'{i}a y Astrof\'{i}sica, UNAM, A.P. 3-72, Xangari, Morelia, 58089, Mexico}
\affil{$^{5}$School of Cosmic Physics, Dublin Institute for Advanced Studies, 31 Fitzwilliam Place, Dublin 2, Ireland}
\affil{$^{6}$Armagh Observatory, College Hill, Armagh BT61 9DG}
\affil{$^{7}$Department of Earth Sciences, National Taiwan Normal University,Taipei, 117 Taiwan}
\affil{$^{8}$Harvard-Smithsonian Center for Astrophysics, 60 Garden Street, Cambridge, MA 02138}

\begin{abstract}
We report JVLA 8-10 GHz ($\lambda$=3.0-3.7 cm) monitoring observations toward the YSO cluster R Coronae Australis 
(R\,CrA), taken in 2012, from March 15 to September 12.
These observations were planned to measure the radio flux variabilities in timescales from 0.5 hours to 
several days, to tens of days, and up to $\sim$200 days.
We found that among the YSOs detectable in individual epochs, in general, the most reddened 
objects in the \textit{Spitzer} observations show the highest mean 3.5 cm Stokes \textit{I} emission, 
and the lowest fractional variabilities on $<$200-day timescales.
The brightest radio flux emitters in our observations are the two reddest sources IRS7W and IRS7E.
In addition, by comparing with observations taken in 1996-1998 and 2005, we found that the radio fluxes of these two 
sources have increased by a factor $\sim$1.5.
The mean 3.5-cm fluxes of the three Class I/II sources IRSI, IRS2, and IRS6 appear to be correlated with their accretion 
rates derived by a previous near infrared line survey.
The weakly accreting Class I/II YSOs, or those in later evolutionary stages, 
present radio flux variability on $<$0.5-hour timescales.
Some YSOs were detected only during occasional flaring events. The source R\,CrA went below our detection 
limit during a few fading events.
\end{abstract}


\clearpage
\section{Introduction }
\label{chap_introduction} 
Young (proto)stars are known to show radio flux variability on various timescales.  
Magnetic reconnections on the (proto)stellar surface can cause non-thermal radio flares in timescales shorter than several 
minutes (Dulk 1985; Bower et al. 2003; Forbrich et al. 2008; Chen et al. 2013; Su et al. 2013), while the interaction of 
the decoupled magnetic fields between the protostars and the disks can result in non-thermal radio flares 
(Shu et al. 1997) on timescales from a few days to as long as the 10-15 days expected from protostellar rotation 
(e.g.  Forbrich et al. 2006, see also Carpenter et al. 2001). 
In addition, accreting young stellar objects (YSOs) can emit thermal radio emission from the regions where the 
magnetohydrodynamic (MHD) wind (Konigl 1982; Pudritz \& Norman 1983, 1986; Shu et al. 1994, 1995) shocks the ambient gas 
(e.g. Rodr{\'{\i}}guez 1997; Rodr{\'{\i}}guez 1999; Anglada 1995; Anglada et al. 1998).
If the accretion rate to the protostar and the mass-loss rate from the protostar are intimately linked as theories suggest 
(Calvet et al. 1993; Shang et al. 2004; see also Chou et al. 2013), then the thermal radio flux is expected to vary 
also on the dynamical timescale of the accretion disk, the timescales of disk instabilities 
(several years; e.g. Zhu et al. 2009), and on the dynamic- and hydrogen recombination timescales of 
the thermal radio jet core (as short as 1-3 months; e.g. Galv{\'a}n-Madrid et al. 2004). 

Radio monitoring observations towards YSOs, planned to resolve the flux variability, spectral indices, and 
polarization percentages, can shed light on discriminating the aforementioned magnetospheric emission mechanisms 
(Forbrich et al. 2011).
In addition, comparison of the radio fluxes between a sample of YSOs occupying a broad range of evolutionary 
stages may provide hints on the evolution of the (proto)stellar magnetosphere on the one-million-year YSO evolutionary 
timescale (e.g. Dzib et al. 2013; AMI Consortium et al. 2012). 
We therefore resumed the multi-epoch 3.5 cm radio observations towards the R Coronae Australis (R\,CrA) cluster 
since 2012 March using the National Radio Astronomy Observatory 
(NRAO)\footnote{The National Radio Astronomy Observatory is a facility of the National Science Foundation operated under 
cooperative agreement by Associated Universities, Inc.} Karl G. Jansky Very Large Array (JVLA).
This target was selected because it has a concentration of early YSOs in a field of a few arcminutes, 
and also due to its proximity ($d\sim$130 pc; for a discussion of the distance, see Neuh{\"a}user \& Forbrich 2008).

\begin{table*}\scriptsize{
\vspace{0cm}
\caption{\footnotesize{The 2012 JVLA observations of 3.5 cm emission.}}
\label{tab_obs}
\vspace{0.3cm}
\hspace{-0.55cm}
\begin{tabular}{lrcccccclcr}\hline\hline
Epoch	&	Time$^{\mbox{\tiny{a}}}$		&	Day$^{\mbox{\tiny{b}}}$	&	Array 	&	$uv$ range$^{\mbox{\tiny{c}}}$	&	Medium 	&
API$^{\mbox{\tiny{d}}}$ 	&	Cloud	$^{\mbox{\tiny{e}}}$	& Synthesized beam  & rms $^{\mbox{\tiny{f}}}$		&	Flux/Pol.\\
			&						&											&		config.				&													&		elevation					&
		rms							&														& $\theta{\mbox{\scriptsize{maj}}}$$\times$$\theta{\mbox{\scriptsize{min}}}$; P.A. & 	noise &	cal.$^{\mbox{\tiny{g}}}$\\
			\hline
			&		(UTC)			&	(day)									&						& (m)												&	(deg)					&
			(deg)							&									&	(arcsec$\times$arcsec, deg) 	& ($\mu$Jy\,beam$^{-1}$)\\
			\hline
1	&	Mar.15	14:21	&	0	&	C	&	26-3387	&	19.1	&	
1.2	&  sky clear	&	8$''$.8$\times$2$''$.4; 177$^{\circ}$	&	19		&	3C286/J2355+4950\\

2	&	Mar.16	14:27	&	1	&	C	&	26-3383	&	19.1	&	
2.0	&  sky clear	&	8$''$.0$\times$2$''$.5; 178$^{\circ}$	&	18		&	3C286/J2355+4950\\

3	&	Mar.17	14:13	&	2	&	C	&	26-3387	&	19.1	&	
1.2	&  10\% covered	&	8$''$.1$\times$2$''$.5; 177$^{\circ}$	&	16		&	3C286/J2355+4950\\

4	&	Mar.17	14:43	&	2	&	C	&	28-3356	&	19.0	&	
1.9	&  10\% covered	&	8$''$.1$\times$2$''$.4; 2.6$^{\circ}$	&	23		&	3C286/J2355+4950\\

5	&	Mar.17	15:13	&	2	&	C	&	30-3314	&	18.5	&	
1.4	&  10\% covered	&	8$''$.4$\times$2$''$.6; 7.2$^{\circ}$	&	18		&	3C48/J2355+4950\\

6	&	Mar.17	15:43	&	2	&	C	&	26-3323	&	17.2	&	
4.8	&  10\% covered	&	9$''$.1$\times$2$''$.6; 14$^{\circ}$	&	21		&	3C48/J2355+4950\\

7	&	Mar.17	16:13	&	2	&	C	&	39-3099	&	15.3	&	
2.5	&  20\% covered	&	9$''$.6$\times$2$''$.7; 20$^{\circ}$	&	24		&	3C48/J2355+4950\\

8	&	Mar.22	13:53	&	7	&	C	&	26-3387	&	19.1	&	
2.2	&  10\% covered	&	9$''$.0$\times$2$''$.4; 178$^{\circ}$	&	20		&	3C286/J2355+4950\\

9	&	Mar.22	14:25	&	7	&	C	&	47-3352	&	19.0	&	
2.7	&  sky clear	&	9$''$.0$\times$2$''$.4; 3.6$^{\circ}$	&	24		&	3C286/J2355+4950\\

10		&	Mar.25	14:14	&	10	&	C	&	28-3352	&	19.0	&	
1.3	&  30\% covered	&	8$''$.3$\times$2$''$.5; 3.4$^{\circ}$	&	22		&	3C286/J2355+4950\\

11		&	Mar.31	13:49	&	16	&	C	&	28-3362	&	19.0	&	
2.4	&  50\% covered	&	8$''$.5$\times$2$''$.3; 2.7$^{\circ}$	&	26		&	3C286/J2355+4950\\

12		&	Apr.02	13:08	&	18	&	C	&	26-3388	&	19.0	&	
6.9	&  10\% covered	&	7$''$.9$\times$2$''$.2; 174$^{\circ}$	&	32		&	3C48/J2355+4950\\

13		&	Jul.28	06:00 &	135	&	B	&	103-11069	&	19.1	&	
7.1	&  10\% covered	&	2$''$.6$\times$1$''$.1; 2.3$^{\circ}$	&	21		&	3C48/J2355+4950\\

14		&	Sep.12	01:45 &	181	&	BnA	&	166-10567	&	17.9	&	
6.5	&  80\% covered	&	2$''$.2$\times$0$''$.96; 166$^{\circ}$	&	34		&	3C286/J2355+4950\\
  \hline
\end{tabular}
}

\vspace{0.1cm}
\footnotesize{Note.--- All epochs were observed using the correlator setting described in Table 
\ref{tab_corr}. 
The pointing center for all epochs of observations is R.A.=19$^{\mbox{\scriptsize{h}}}$01$^{\mbox{\scriptsize{m}}}$48.000$^{\mbox{\scriptsize{s}}}$ (J2000), Decl.=-36$^{\circ}$57$'$59$''$.0 (J2000).
}
\par
\scriptsize{
\begin{itemize}
\item[$^{\mbox{\scriptsize{a}}}$] All epochs are observed in 2012. The observations (including calibrations) started 15 
minutes before the time noted here, and ended 15 minutes after it. The first $\sim$10 minutes in each epoch were used for 
taking dummy observing scans as required by the system.\vspace{-0.15cm}
\item[$^{\mbox{\scriptsize{b}}}$] The relative day to the first epoch. \vspace{-0.15cm}
\item[$^{\mbox{\scriptsize{c}}}$] From the minimum to the maximum of the baseline projected lengths. We present it in units of meters rather than kilo-wavelengths because of the large range of observing frequencies.  \vspace{-0.15cm}
\item[$^{\mbox{\scriptsize{d}}}$] The values of the atmospheric phase interferometer quoted from the observing log.  \vspace{-0.15cm}
\item[$^{\mbox{\scriptsize{e}}}$] The sky condition commented by the JVLA operator. \vspace{-0.15cm}
\item[$^{\mbox{\scriptsize{f}}}$] Measured at the center of the IF1 Stokes \textit{I} image generated utilizing the 1 GHz total bandwidth (centered at the sky frequency $\nu$=8.5 GHz). \vspace{-0.15cm}
\item[$^{\mbox{\scriptsize{g}}}$] The observed quasar for absolute flux and polarization calibrations. 
\end{itemize}
\vspace{-0.1cm}}
\end{table*}

The R\,CrA cluster is one of the nearest young, dense (i.e. $>$25 Class 0-II YSOs pc$^{-2}$, see Myers 2009)  clusters 
which remains embedded in the natal molecular cloud. 
The previous optical and near-infrared (OIR) observations (Wilking et al. 1985, 1992; L{\'o}pez Mart{\'{\i}} et al. 2005; 
Haas et al. 2008; Peterson 2011) found that the majority of the objects younger than Class II are located in the central 
$r\sim$0.1 pc ($\sim$2.4$'$) gas concentration (Loren 1979; Loren et al. 1983; Harju et al. 1993; Henning et al. 1994; 
Andreazza \& Vilas-Boas 1996; Anderson et al. 1997a, 1997b; Chini et al. 2003; Groppi et al. 2004).
High angular resolution mapping observations and molecular line surveys further confirmed abundant protostellar 
cores in this region (Nutter et al. 2005; Sch{\"o}ier et 
al. 2006; Lindberg \& J{\o}rgensen 2012; Watanabe et al. 2012; Sicilia-Aguilar et al. 2011, 2013) and even found prestellar 
core candidates (Groppi et al. 2007; Chen \& Arce 2010).
The instantaneous accretion rates of several YSOs in this field have been constrained by a near infrared line survey in 2002 July 12 and 13 (Nisini et al. 2005).

In the radio part of the spectrum, VLA 6-cm observations in 1985 resolved 11 radio emission sources at $>$4.5 $\sigma$ 
significance  (Brown 1987).
Similar results were also given by the Australia Telescope Compact Array (ATCA) 3-cm, 6-cm, and 20-cm observations in 
1998 and 2000 (Miettinen et al. 2008). 
Early VLA and Australia Telescope (AT) observations between 1985 and 1993 have revealed radio flux variability  
at 6-cm wavelength (Suters 1996). 
More extensive, deep 3.5-cm observations were performed with the VLA in 1996-1998 
(Feigelson et al. 1998; Forbrich et al. 2006), and in 2005 (Forbrich et al. 2007; Choi et al. 2008, 2009). 
Due to limited sensitivity, those previous radio observations generally have an on-source integration time of 
several hours to achieve an adequate significance for detections.

Thanks to the improved sensitivity of the JVLA, investigation of the radio flux variability in the 
R\,CrA cluster on $\ll$1-hour timescales is now feasible. 
Comparing our follow up JVLA observations in 2012 with the previous VLA observations in 1997, 1998, and 2005, 
will provide time baselines for examining the radio flux variability in the decadal accretion-disk evolutionary timescale. 
We introduce the details of our 2012 observations in Section \ref{chap_obs} .
The results of our observations are given in Section \ref{chap_result}.
The statistics of our observational results are presented in Section \ref{sub_biweight}.
By incorporating earlier observational results, in Section \ref{sub_structure} we present the $\sim$15-year timescale 
Stokes \textit{I} radio flux variability.
A preliminary interpretation of our observational results is given in Section \ref{sub_interpretation}.
Section \ref{chap_summary} summarizes the main results of this paper.

\begin{table}\scriptsize{
\vspace{0.2cm}
\caption{\footnotesize{The correlator setup of the 2012 JVLA observations.}}
\label{tab_corr}
\vspace{0.1cm}
\hspace{-0.1cm}
\begin{tabular}{ccccc}\hline\hline
IF	&	Spw ID$^{\mbox{\tiny{a}}}$	&	Central frequency$^{\mbox{\tiny{b}}}$	&	Bandwidth	& \# of spectral channels	\\
	&							&	(MHz)													&	(MHz)			&						\\\hline
\multirow{8}{*}{1}	&	0	&	8051		&	128			&		128		\\
						&	1	&	8179		&	128			&		128		\\
						&	2	&	8307		&	128			&		128		\\
						&	3	&	8435		&	128			&		128		\\
						&	4	&	8563		&	128			&		128		\\
						&	5	&	8691		&	128			&		128		\\
						&	6	&	8819		&	128			&		128		\\
						&	7	&	8947		&	128			&		128		\\\hline
\multirow{8}{*}{2}	&	8	&	9051		&	128			&		128		\\
						&	9	&	9179		&	128			&		128		\\
						&	10	&	9307		&	128			&		128		\\
						&	11	&	9435		&	128			&		128		\\
						&	12	&	9563		&	128			&		128		\\
						&	13	&	9691		&	128			&		128		\\
						&	14	&	9819		&	128			&		128		\\
						&	15	&	9947		&	128			&		128		\\\hline

\end{tabular}
}

\vspace{0.1cm}
\footnotesize{Note.--- 
There is no doppler tracking in our observations. 
}
\par
\scriptsize{
\begin{itemize}
\item[$^{\mbox{\scriptsize{a}}}$] The ID of the observed spectral windows. \vspace{-0.15cm}
\item[$^{\mbox{\scriptsize{b}}}$] The sky frequency at the center of the spectral window. The spectral windows 
spw 10 and 11 {\bf often had} strong radio frequency interference (RFI) and thus were flagged out for 
all epochs of observations.
\end{itemize}
\vspace{0cm}}
\end{table}


\begin{figure*}
\vspace{-3cm}
\hspace{-1.3cm}
\begin{tabular}{p{9.2cm} p{9.2cm} }
\includegraphics[width=11.7cm]{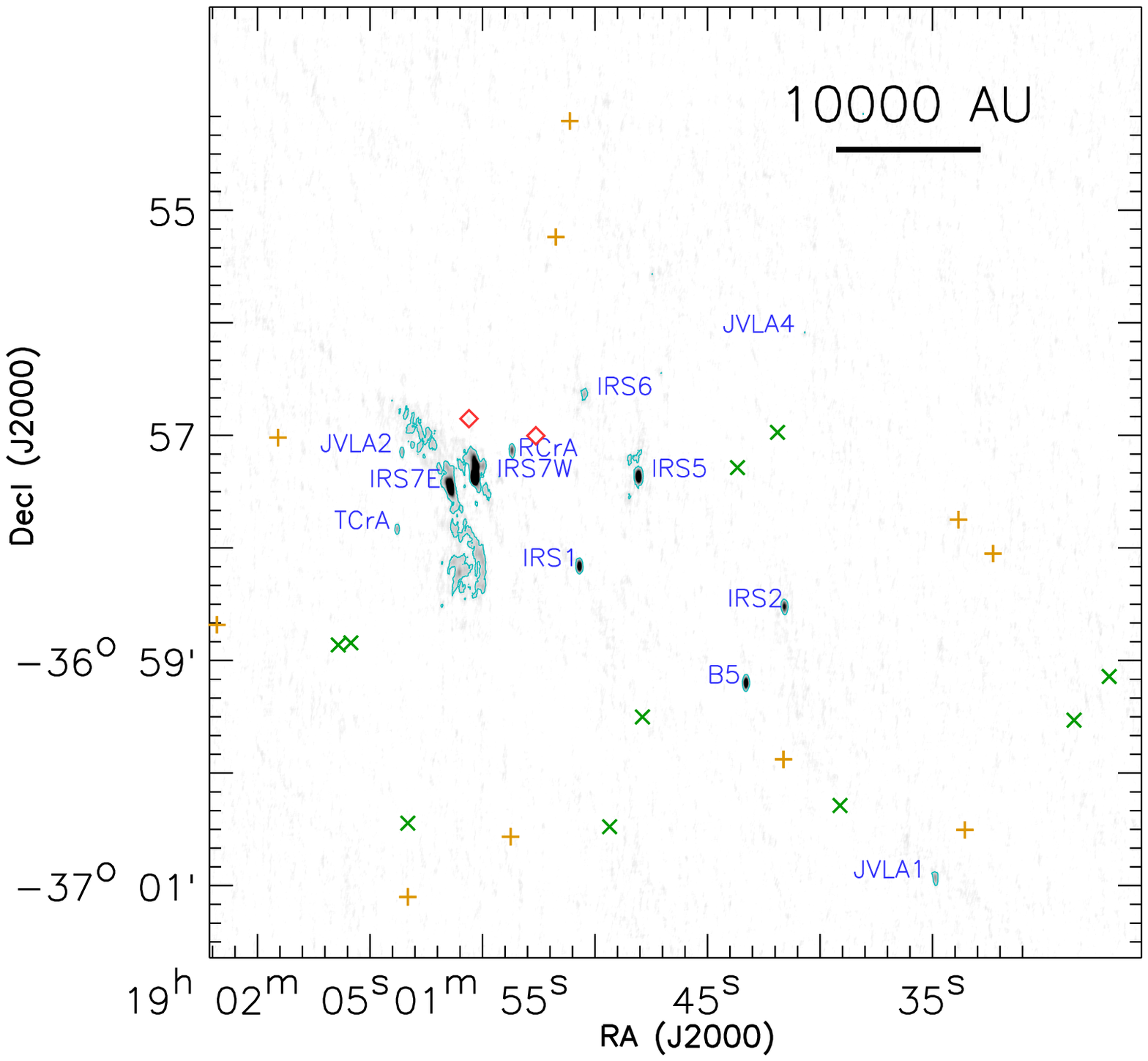} & \includegraphics[width=11.7cm]{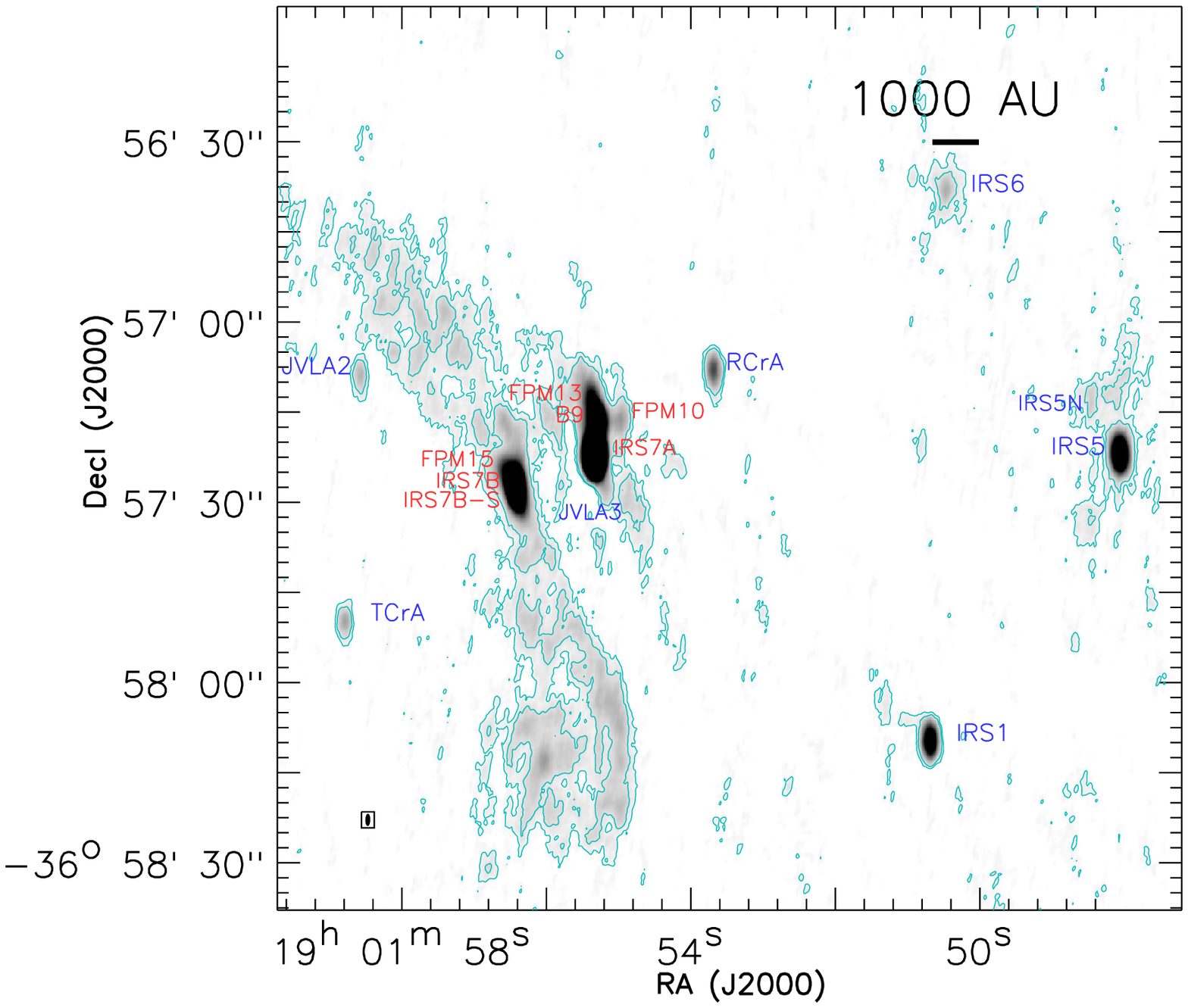}	\\
\end{tabular}
\caption{\footnotesize{
The 3.5-cm radio image of the R\,CrA YSO cluster (gray scale). The right panel zooms into the sub-field around 
the groups of compact radio sources IRS7E and IRS7W (Table \ref{tab_sources}).   
This image is generated using Briggs weighting with Robust=1, incorporating all IF1 data described in 
Tables \ref{tab_obs} and \ref{tab_corr}. 
The $\theta_{\mbox{\scriptsize{maj}}}$$\times$$\theta_{\mbox{\scriptsize{min}}}$=4$''$.3$\times$2$''$.0, P.A.=-179$^{\circ}$ synthesized beam is shown in bottom left of the right panel. 
The rms noise level is 8.5 $\mu$Jy\,beam$^{-1}$. 
Contours in the left and right panels are [5$\sigma$] and [2.5$\sigma$, 5$\sigma$], respectively. 
To avoid noisier edges, the images presented are not yet corrected for primary beam attenuation. 
The annotated sources and the primary beam attenuation at their location can be found in Tables 
\ref{tab_sources} and \ref{tab_PB}. 
The scale bars in both panels are drawn assuming a distance of 130 pc (see Deller et al. 2013 and references 
therein).
The red diamonds, orange crosses, and green crosses mark the locations of the Class I, Class II and 
flat SED, and Class III YSOs (Peterson et al. 2011) which were not detected in our JVLA observations, 
respectively. 
}
\vspace{0.3cm}
}
\label{fig_img}
\end{figure*}

\section{Observations and Data Reduction} 
\label{chap_obs}

We performed 16 epochs of filler-mode observations towards the R Coronae Australis region using the JVLA C, B, and BnA array 
configurations in 2012 from March to September. 
The pointing center for all epochs is R.A.=19$^{\mbox{\scriptsize{h}}}$01$^{\mbox{\scriptsize{m}}}$48.000$^{\mbox{\scriptsize{s}}}$ (J2000), Decl.=-36$^{\circ}$57$'$59$''$.0 (J2000).
Each observation epoch has an overall duration of 30 minutes, and contains two $\sim$220 
seconds$\footnote{The exact on-source time slightly varies among epochs of observations because of the 
differences in antenna slewing time.}$ on-source scans (separated by $\sim$50 seconds). 
This relatively short calibration cycle as compared with the typically longer than 20 minutes calibration duty cycles, 
helps compensate out the relatively large atmospheric effects for the low elevation target source 
(also mentioned in Forbrich et al. 2006). 
We lost one epoch of observations on 2012 March 25 because of missing the calibration data; and we lost another 
epoch of observations in 2012 September 13 because the weather conditions were too poor to allow robust 
antenna-based gain calibrations.
The details of the remaining observations are listed in Table \ref{tab_obs}.
The correlator setup of our observations can be found in Table \ref{tab_corr}.
The total bandwidth after combining the 16 spectral windows in the 2 independently tunable intermediate frequencies (IFs) 
is 2 GHz (Table \ref{tab_corr}).
We centered IF1 at a sky frequency of 8.5 GHz to enable comparison with the extensive earlier VLA observations 
at the same frequency (Feigelson et al. 1998; Forbrich et al. 2006; Forbrich et al. 2007; Choi et al. 2008, 2009).
The IF2 was tuned to complete a continuous 2 GHz total frequency coverage, and also to minimize 
differences in the primary beam coverage between the lowest and the highest frequency spectral windows.
The expected root-mean-square (rms) noise level for individual images of a 128 MHz spectral window is 
$\sim$50 $\mu$Jy\,beam$^{-1}$ in each epoch. However, the noise may be degraded depending on the data flagging 
and the unresolved radio frequency interference (RFI).

The data were calibrated using the Common Astronomy Software Applications (CASA)\footnote{http://casa.nrao.edu} package release 3.4.0.
Each epoch of observations listed in Table \ref{tab_obs} was further independently phase self-calibrated to minimize the decoherence of fluxes.
By the time these data were calibrated, the JVLA data format did not include weights, and the CASA task \texttt{statwgt} for reweighting visibilities was still experimental.
Because of being unable to calibrate the weightings, we were not able to correctly combine the data in spectral windows with very different noise levels. 
Therefore, in this manuscript, we only present the data in spectral windows 0-8, which have the lowest noise levels; 
and we only jointly imaged the data in IF1. 
Including the other spectral windows in the joint imaging either does not change the results, or may increase the noise 
level.
However, the weighting issue does not affect the studies presented in this paper.
We expect this weighting issue to be resolved in the future, which will allow us to reprocess these data.

\begin{table*}\scriptsize{
\vspace{0cm}
\caption{\footnotesize{The 2-dimensional Gaussian components for initializing the source fits.}}
\label{tab_sources}
\vspace{0.3cm}
\hspace{-0.2cm}
\begin{tabular}{lcccccccc}\hline\hline
Source name	&		&		R.A.		&	Decl.		&		Major axis FWHM$^{\mbox{\tiny{a}}}$		&		Minor axis FWHM$^{\mbox{\tiny{a}}}$
		&		P.A.	&	Flux	&	P11 Classification$^{\mbox{\tiny{b}}}$ \\	
					&		&		(J2000)	&	(J2000)	&		(arcsec)					&		(arcsec)	&		(deg)		&		(mJy)		&			\\\hline

IRS7B	&	\multirow{3}{*}{IRS7E}	&	19:01:56.422		&		-36:57:27.6		&		2.88		&		1.24		&		5.0		&		1.46		&		\multirow{3}{*}{Class I}	\\

FPM15	&									&	19:01:56.476		&		-36:57:25.6		&		5.60		&		2.00		&		18.9		&		0.84		&										\\

IRS7B-S &									&	19:01:56.326		&		-36:57:30.8		&		3.49		&		1.22		&		178.1	&		0.25		&										  \\\hline

IRS7A	& \multirow{4}{*}{IRS7W}	&	19:01:55.325		&		-36:57:22.1		&		2.74		&		1.31		&		175.9		&		6.63		&		\multirow{4}{*}{Class I} \\

B9			&									&	19:01:55.291		&		-36:57:16.6		&		2.69		&		1.31		&		176.0		&		1.19		&		\\

FPM13	&									&	19:01:55.375		&		-36:57:13.0		&		5.56		&		1.71		&		178.3		&		0.66		&		\\

FPM10	&									&	19:01:54.974		&		-36:57:16.0		&		4.80		&		2.36		&		178.5		&		0.26		&		\\\hline

JVLA3 (CXO\,34)	&					&	19:01:55.793		&		-36:57:27.1		&		2.10			&	0.96		&		172.0		&		0.060	&		Class I		\\

JVLA2 (WMB55)		&					&	19:01:58.561		&		-36:57:08.6		&		2.70		&		1.20		&		178.0		&		0.10		&		Class I		\\
		
IRS5N	&					&	19:01:48.484						&		-36:57:14.8		&		2.10			&	0.69				&	166.0	&		0.049			&		Class I		\\

IRS5		&									&	19:01:48.061		&		-36:57:22.0		&		3.05		&		1.42		&		0.9			&		1.09		&		Class I		\\

IRS1		&									&	19:01:50.685		&		-36:58:09.7		&		2.98		&		1.21		&		179.3		&		0.64		&		Class I		\\

JVLA4 (Haas 4) 		&					&	19:01:40.667		&		-36:56:05.2		&		2.10			&	0.96		&		169.0		&		0.084	&		Flat SED		\\

IRS2		&									&	19:01:41.579		&		-36:58:31.3		&		2.88		&		1.20		&		178.4		&		0.37		&		Class I		\\

IRS6		&									&	19:01:50.484		&		-36:56:38.3		&		3.83		&		1.26		&		167.0		&		0.12		&		Class II		\\

T\,CrA		&									&	19:01:58.784		&		-36:57:49.7		&		3.18		&		1.70		&		176.0		&		0.18		&		Class II		\\

JVLA1 (CrA\,PMS1)		&					&	19:01:34.858		&		-37:00:55.7		&		2.69		&		1.11		&		178.2		&		0.13		&		Class III		\\

R\,CrA 	&									&	19:01:53.686		&		-36:57:08.0		&		3.49		&		1.22		&		178.1		&		0.28		&		Class III		\\

B5 &					&	19:01:43.283		&		-36:59:12.0		&		2.71		&		1.14		&		174.9		&		0.68		&		Galaxy		\\

\hline

\end{tabular}
}

\vspace{0.45cm}
\footnotesize{Note.--- This target list is generated by fitting the compact sources in the deep Briggs Robust=0 
weighted image, incorporating all IF1 data described in Tables \ref{tab_obs} and \ref{tab_corr}. 
IRS7E resolved at higher angular resolotion into IRS7B, FPM15, and IRS7B-S. IRS7W resolves into
IRS7A, B9, FPM10, and FPM13.
The 1$\sigma$ rms noise levels at the individual locations of these sources are 
$\sim$8.5 $\mu$Jy\,beam$^{-1}$ divided by the primary beam attenuation factors listed in Table \ref{tab_PB}.
}
\par
\scriptsize{
\begin{itemize}
\item[$^{\mbox{\scriptsize{a}}}$] The listed values of FWHM are not yet deconvolved from the 
$\theta_{\mbox{\scriptsize{maj}}}$$\times$$\theta_{\mbox{\scriptsize{min}}}$=2$''$.7$\times$1$''$.11 (P.A.=178$^{\circ}$) 
synthesized beam. Several of the listed sources are consistent with point sources at our angular resolution, 
thus cannot be deconvolved. \vspace{-0.15cm}
\item[$^{\mbox{\scriptsize{b}}}$] YSO classification quoted from Peterson et al. (2011), except for 
the extragalactic source B5.
\end{itemize}
\vspace{0.4cm}}
\end{table*}

\begin{table}\scriptsize{
\vspace{0cm}
\caption{\footnotesize{The averaged primary beam attenuation factor for individual sources.}}
\label{tab_PB}
\vspace{0.3cm}
\hspace{1.2cm}
\begin{tabular}{lcccccccc}\hline\hline
Source name		&			Spw\,1		&		Spw\,8		&		IF1$^{\mbox{\tiny{a}}}$		\\
\hline
IRS7E						&	0.76			&		0.71			&		0.74										\\
IRS7W						&	0.79			&		0.75			&		0.78										\\
JVLA3 (CXO\,34)		&	0.79			&		0.74			&		0.77										\\
JVLA2 (WMB55)			&	0.62			&		0.56			&		0.60										\\
IRS5N						&	0.94			&		0.94			&		0.95										\\
IRS5							&	0.97			&		0.96			&		0.96										\\
IRS1							&	0.97			&		0.97			&		0.97										\\
JVLA4 (Haas\,4)			&	0.59			&		0.52			&		0.56										\\
IRS2							&	0.84			&		0.81			&		0.83										\\
IRS6							&	0.83			&		0.80			&		0.82										\\
T\,CrA						&	0.65			&		0.59			&		0.63										\\
JVLA1 (CrA\,PMS1)		&	0.20			&		0.13			&		0.18										\\
R\,CrA						&	0.84			&		0.80			&		0.83										\\
B5								&	0.81			&		0.77			&		0.80										\\
\hline
\end{tabular}
}

\vspace{0.1cm}
\footnotesize{Note.--- The columns Spw\,1 and Spw\,8 are the primary beam attenuation factors for images generated using the data in spectral window 1 and in spectral window 8, respectively. 
}
\par
\scriptsize{
\begin{itemize}
\item[$^{\mbox{\scriptsize{a}}}$] The averaged primary beam attenuation factors while incorporating all spectral windows in IF1.
\end{itemize}
\vspace{0.1cm}}
\end{table}

We performed the naturally weighted imaging using the CASA task \texttt{clean}. 
The image size is 3600 pixels in each dimension, and the pixel size is 0.2$''$.
A few of the sources in the R\,CrA field are known to be associated with extended radio emission (e.g., Miettinen et al. 2008).
We therefore implemented a lower cut of the visibility $uv$ distances ($\sqrt{u^2 + \nu^2}$) of 4.4 k$\lambda$, 
which is comparable to the shortest baseline of the BnA array observations (Table \ref{tab_obs}), 
to all epochs of data before imaging.
This ensures that the extended emission does not bias the measurements of the flux variations.
We found that most of the emission comes from the compact components after implementing the 4.4 k$\lambda$ cut.
The flux measurements are not sensitive to the $\sqrt{u^2 + \nu^2}$ cut when it is longer than 4.4 k$\lambda$.
Cutting even at much larger $uv$ distances (e.g. 45 k$\lambda$) does not fundamentally change our measurements, 
however, it significantly degrades the sensitivity and the synthesized beam shapes of the C array observations.
The rms noise levels achieved after combining the spectral windows 0-7 are given in Table \ref{tab_obs}. 
We note that these observations are sensitive to events at the 0.6-0.9 mJy level, such as the radio-jet knot eruption 
reported by Pech et al. (2010) in IRAS 16293-2422.

\vspace{-0.1cm}
\section{Results}
\label{chap_result}

\subsection{The Compact Sources in the 3.5 cm Stokes \textit{I} Image}
\label{sub_stokesI}
To yield a deep radio image, we combined and jointly imaged the phase self-calibrated IF1 data (Table \ref{tab_corr}) from all 14 epochs of observations listed in Table \ref{tab_obs}.
The Briggs Robust=1 weighted combined image without the implementation of the $>$4.4 k$\lambda$ $uv$ distance limit (Section \ref{chap_obs}) is shown in Figure \ref{fig_img}.
The compact radio sources were registered by performing 2-dimensional Gaussian fittings on the Briggs Robust=0 weighted combined image, using the CASA task \texttt{imfit}.
Since the Gaussian fitting is fundamentally ambiguous (e.g., not necessarily converges to a unique solution), we implemented the minimal possible number of Gaussian components which can recover the emission of the compact sources well.  
The Gaussian components are listed in Table \ref{tab_sources}.
The primary beam attenuation of the fluxes was corrected only after the 2-dimensional Gaussian fittings to avoid 
confusion by the noise.
The primary beam attenuation factors for the individual of sources are given in Table \ref{tab_PB}.
Hereafter, we refer to the group of sources IRS7B, FPM15, and IRS7B-S as IRS7E, and to the group of sources IRS7A, B9, FPM13, 
and FPM10 as IRS7W, because  they are not resolved in every epoch of our JVLA C array observations.
The source IRS5 is known to be binary (its components are known as IRS5a, IRS5b, see Chen \& Graham 1993; Choi et al. 2008; Deller et al. 2013), 
however, they cannot be separated given our angular resolution.
The radio flux variability of the individual components in these groups will not be independently discussed in this 
manuscript (Section \ref{sub_Ivar}).
From high angular resolution 7 mm continuum images (Choi \& Tatematsu 2004), the group IRS7W is likely to be a cluster of young (proto)stars with associated thermal jet knots.
The components FPM15 and IRS7B-S may trace discrete knots in the extended bipolar radio jet emanating from the young stellar object IRS7B (Forbrich et al. 2006; Choi et al. 2008).
Alternatively, they may be tracing the base of this outflow. 
A future search for proper motions may clarify the nature of these components.

We found that the sources IRS7E, IRS7W, IRS5, and IRS6 are associated with extended radio emission (Figure \ref{fig_img}).
The properties of the extended emission will be addressed in a forthcoming paper incorporating the follow up 
observations in the more compact JVLA array configuration, and the observations at higher frequencies.
The radio source JVLA1 is detected for the first time, and is likely the known Class III object CrA\,PMS\,1.
By cross comparing with the YSO source catalog from the previous infrared surveys (Peterson et al. 2011), 
we also claim the new detections of the additional two faint sources JVLA3 and JVLA4,  which are likely the 
Class I object CXO\,34, and the flat spectrum object Haas\,4, respectively.
JVLA3 can be isolated from the extended emission after the $>$4.4 k$\lambda$ $uv$ distance limit is implemented.
In the image incorporating data from all epochs of observations, the source JVLA2 can be isolated from the north-east extended radio lobe emanated from IRS7E.
It may be the weakly detected radio source WMB55 reported in Choi et al. (2008) (see also Wilking et al. 1997), which is associated with the submillimeter core SMM2 (Groppi et al. 2007).

The radio source B5 which was previously proposed to be a brown dwarf candidate (Feigelson et al. 1998), 
is now confirmed to be extragalactic (Jan Forbrich, private communication), thus will be omitted 
in the following discussion. 

\begin{figure}
\vspace{-0.7cm}
\hspace{-0.7cm}
\includegraphics[width=10.5cm]{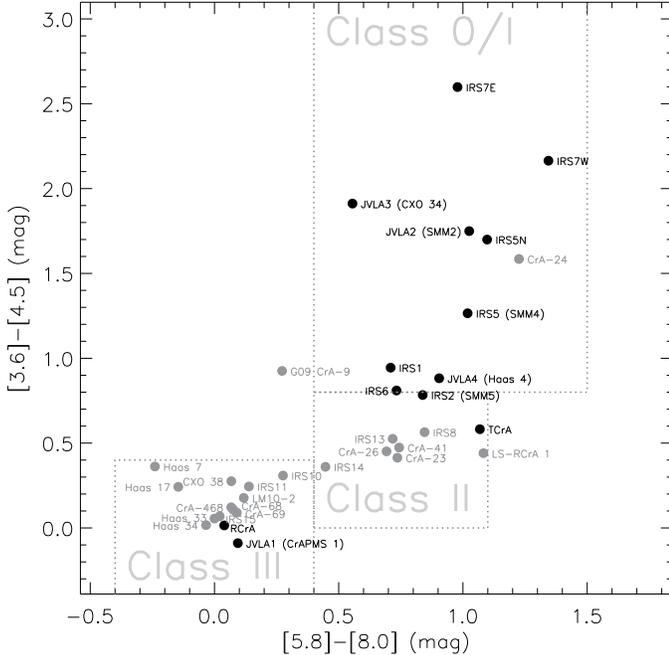}
\vspace{-1cm}
\caption{\footnotesize{
The \textit{Spitzer} color-color diagram of the detected YSO radio sources (black). 
The YSOs located within the JVLA primary-beam attenuation contour at 0.1 of the peak, 
but that were not detected in radio emission, are plotted in gray symbols. 
The plotted data are quoted from Peterson et al. (2011). 
The dashed loci for YSO classification are drawn based on Allen et al. (2004) and Lee et al. (2006).
The source R\,CrA is in fact a Herbig Ae star (Kraus et al. 2009, and references herein).
}
}
\label{fig_spitzer}
\end{figure}

To provide a sense of the evolutionary stages of the detected YSOs, we quote the Peterson et al. (2011) classification 
of YSOs in Table \ref{tab_sources}, and show the \textit{Spitzer} color-color diagram in Figure \ref{fig_spitzer}.
The \textit{Spitzer} fluxes in the color-color diagram are also from Peterson et al. (2011).
The \textit{Spitzer} fluxes of the well known Class I YSO candidate IRS9 
(Forbrich \& Preibisch 2007; Peterson et al. 2011) cannot be measured because it is located 
too close to the bright source R\,CrA.
We do not detect  3.5-cm radio emission from IRS9 either and will omit this source from the following discussion.

The YSOs at earlier evolutionary stages should in general be redder and will appear at the 
top right of the \textit{Spitzer} color-color diagram.
Contamination and short-period infrared variations of the YSOs may cause uncertainties in the \textit{Spitzer} colors.
We note that although the loci in the \textit{Spitzer} color-color diagram help to divide the 
sample into the conventional Class 0-III evolutionary stages, the actual evolutionary tracks of the YSOs may be more continuous.
For sources located very near a boundary, for example, IRS2 and IRS6, their classification as Class I or II is not intrinsically important. 
We also note that the source R\,CrA is in fact a Herbig Ae star (Peterson et al. 2011, and references therein).
By comparing Figure \ref{fig_img} with Figure \ref{fig_spitzer}, we conclude that the non-detections of 
some Class II and Class III YSOs are not due to primary-beam attenuation. 
It is more likely that the majority of these sources were fainter than our sensitivity limit at all epochs.

Figure \ref{fig_spitzer} shows that while practically all Class 0/I sources are detected in the radio, 
sources in the Classes II/III are only rarely detected. 
A possible explanation is that in the Class 0/I sources, we are detecting free-free emission from an ionized outflow 
that is systematically present in this type of objects. 
In contrast, in the Class II/III sources, we are probably detecting gyrosynchrotron emission from active magnetospheres (see also Gibb 1999). 
This is a time-variable process that is not necessarily present in all Class II/III stars 
(more discussion in Section \ref{sub_interpretation}). 
In addition, the left panel of Figure \ref{fig_img} shows a clear spatial differentiation of the Clsss 0/I and 
Class II/III sources. 
While Class 0/I sources are confined to the inner part of the cluster, in a region of about 2$'$ in extent, the 
Class II/III sources extend over 4$'$-5$'$. 
This may suggest that star formation did not take place simultaneously in all the cluster, but that it propagated 
inwards with time.
Alternatively, it may be explained by the fact that the Class II and III objects are old enough and have had time 
to diffuse (a star moving at $\sim$1 km\,s$^{-1}$ travels 1 pc in 1 Myr).

\begin{figure}
\begin{tabular}{p{9cm} }
	\\
\end{tabular}

\vspace{-2.6cm}

\hspace{-2cm}
\begin{tabular}{p{9cm} }
\includegraphics[width=11.5cm]{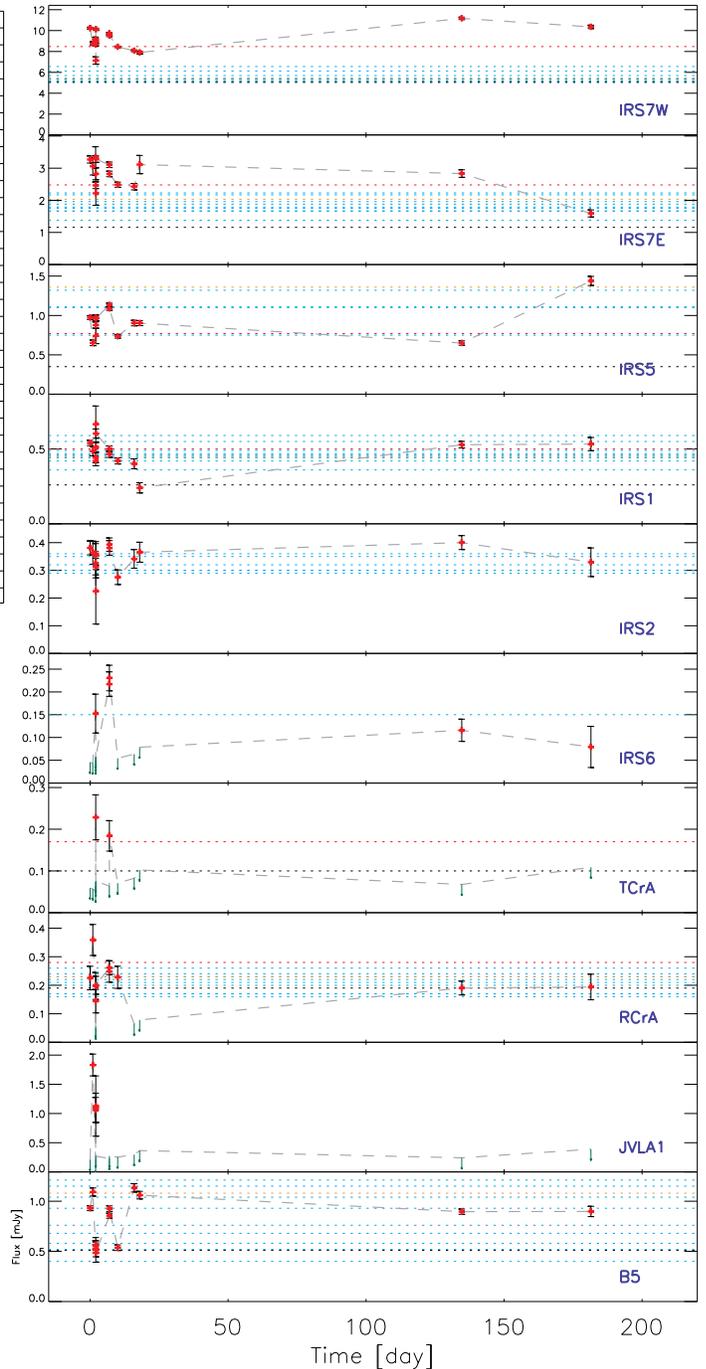} 	\\
\end{tabular}
\vspace{-2.7cm}
\caption{\footnotesize{
The 3.5 cm Stokes \textit{I} fluxes of the detected sources. The source JVLA2 is too faint to be robustly 
detected in individual epochs, thus cannot be plotted. 
The red symbols connected with gray dashed lines show the 2012 observations listed in Table \ref{tab_obs}. 
For particular 2012 epochs in which the sources cannot be detected, the 2$\sigma$ upper limits are given 
in green downward arrows. 
Otherwise, we show the $\pm$1 $\sigma$ uncertainties with black error bars. 
In addition, we quote the 3.5-cm Stokes \textit{I} fluxes observed in 1996 December (Choi et al. 2008), 1997 January (Feigelson et al. 1998), 1998 from June to October (Forbrich 2006), and 2005 February (Choi et al. 2008), in black dotted lines, orange dotted lines, cyan dotted lines, and red dotted lines, respectively. 
The 1998 flux of IRS6 (and of FPM10, FPM13, FPM15, which areincorporated in IRS7W and IRS7E), is averaged 
from all 9 epochs of observations in 1998 because of the faintness of the source.
The source T\,CrA was not detected in 1998 observations. 
The sources JVLA1 was not detected in any of the previous observations.
}
}
\label{fig_flux1}
\end{figure}

\begin{figure*}
\vspace{-2cm}
\hspace{-1.7cm}
\begin{tabular}{p{9cm} p{9cm} }
\includegraphics[width=11.5cm]{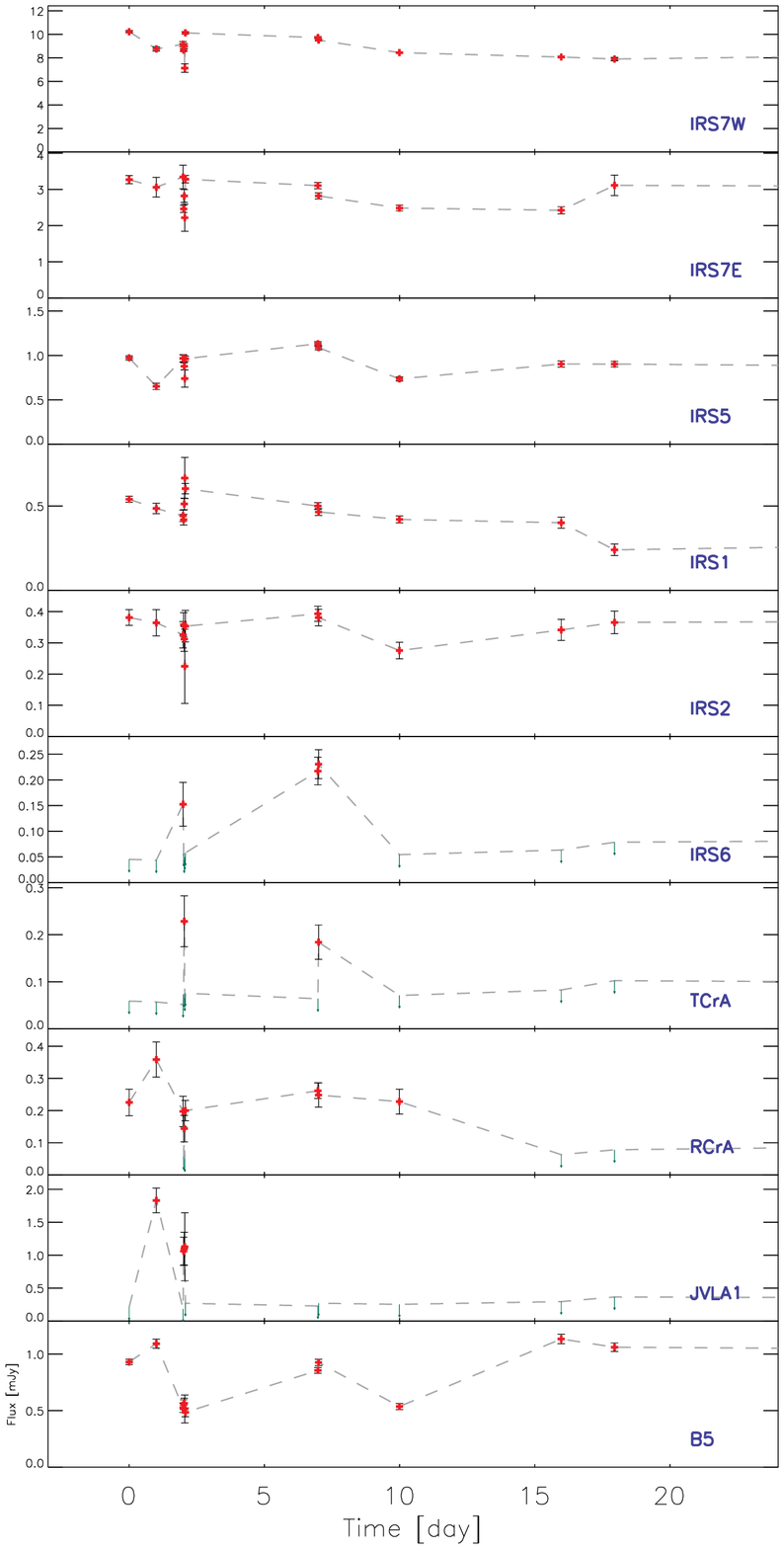} & \includegraphics[width=11.5cm]{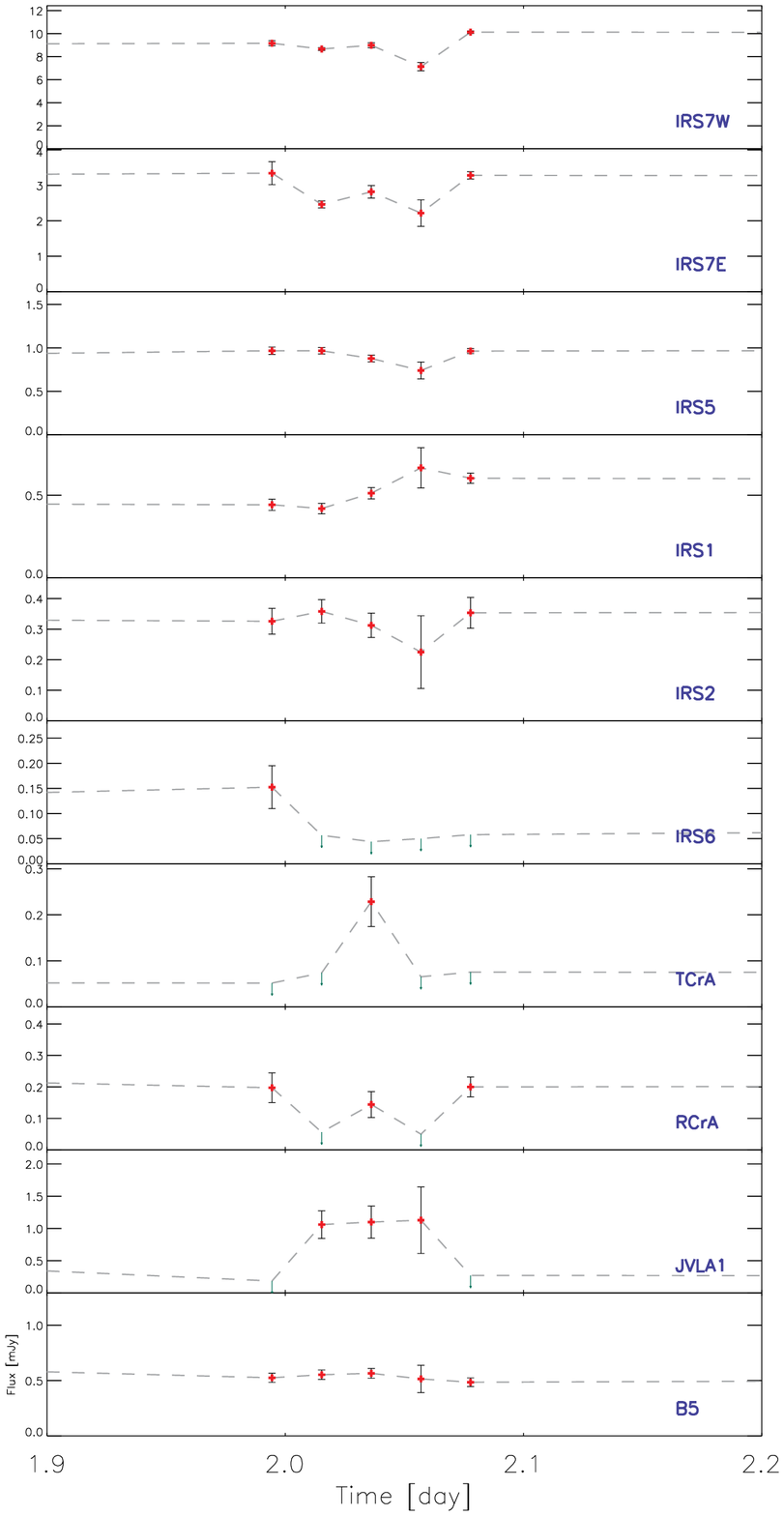}	\\
\end{tabular}
\vspace{-2.7cm}
\caption{\footnotesize{
Similar to Figure \ref{fig_flux1}. The left panel zooms into the time periods of the C-array 
observations listed in Table \ref{tab_obs}. The right panels zooms in further to show only the 5 epochs 
taken on 2012 March 17.
}
}
\label{fig_flux2}
\vspace{0.4cm}
\end{figure*}

\subsection{The 3.5 cm Stokes \textit{I} Flux Variabilities.}
\label{sub_Ivar}
To analyze the 3.5-cm Stokes \textit{I} flux variabilities, images with low noise and minimal phase decoherence are required. 
For each epoch of observations, we therefore jointly imaged the $\sqrt{u^2 + \nu^2} > $4.4 k$\lambda$ phase 
self-calibrated data in IF1 (Table \ref{tab_corr}).
We perform 2-dimensional Gaussian fits to these broad band images to obtain the fluxes.
The derived Gaussian components in Table \ref{tab_sources} were used to initialize the Gaussian fits 
for all epochs. 
The residual noise level, as well as the shapes of the Gaussian models and the residuals, were inspected to 
verify the convergence of the fits. 
The errors from the Gaussian fits were obtained by taking the maximum of the two estimates 
described in the AIPS++ Note 
244\footnote{{\scriptsize http://www.astron.nl/casacore/trunk/casacore/doc/notes/224.html}}, and in Condon et al. 
(1998), as well as Richards et al. (1999).
These two methods are based on the signal to noise ratio of the fitted Gaussian component, and the goodness 
of fit, respectively.
For most of the epochs, the positions of each Gaussian component only need to be shifted by $<$2 pixels (0.4$''$) 
relative to the initial model, which is not significant as compared with the synthesized beam sizes 
(Table \ref{tab_obs}).
The images of Epoch 6, 7, and 12 required to shift the Gaussian component relative to the initial model, 
by up to 20\% of the synthesized beam full width half maximum (FWHM). This probably results from a 
combination of noise and spatial drifts of the images due to the phase self-calibration.

Alternative methods to obtain the fluxes are summing the fluxes within box regions enclosing the sources 
(e.g. Feigelson et al. 1998), or summing the fluxes in regions (partially) defined by contours at certain 
significance levels (e.g. 2$\sigma$, Choi et al. 2008).
We did not use the former method because it is hard to uniformly define the box regions given the variations 
of the synthesized beam sizes and position angles in our 2012 observations (Table \ref{tab_obs}).
The latter method is potentially biased in observations with high noise levels.
In our observations, the differences of the measured fluxes with all mentioned methods are typically less than 10\% for IRS7W and IRS7E, and are much smaller for point sources. 
This systematic effect is smaller than the intrinsic flux variabilities for most of the sources (Section \ref{sub_biweight}). 
However, these methods are subject to different errors (i.e., the error bars can be different). 

The measured broad band 3.5-cm Stokes \textit{I} fluxes are shown in Figure \ref{fig_flux1}.
We also quote the 3.5-cm fluxes in earlier VLA A-array observations on 1996 December 29 (Choi et al. 2008), 
BnA-array observations on 1997 January 19 and 20 (Feigelson et al. 1998), BnA array observations on 1998 June 27 
(Forbrich et al. 2006), 8 epochs of B-array observations on 1998, from July 19 to October 13 
(Forbrich et al. 2006), and BnA-array observations on 2005 February 03 (Choi et al. 2008).
We note that the radio flux of IRS5 flared to up to $\sim$3.3 mJy in some epochs in 1998, 
which exceeds the plotted range. 
The flux of IRS2 on 1997 January 19-20 is 0.67 mJy, which also exceeds the plotted range. 
These large variations of bright sources will be addressed in Section \ref{chap_discussion}.
Figure \ref{fig_flux2} zooms in to better present the observations from 2012 March 15 to April 02, 
and the consecutive 5 epochs of observations on 2012 March 17. 

In our JVLA field of view (Figure \ref{fig_img}), three Class 0/I sources (IRS7E, IRS7W, IRS5), 
and one source in between the Class I and Class II stages (IRS6) are associated with diffuse Stokes \textit{I} 
emission.
However, two of these sources (i.e. IRS7W and IRS5) show higher 3.6-cm fluxes in the more 
extended B-array and BnA-array measurements, which can only be explained by flux variations on the 
unresolved spatial scales.
The source IRS 7E shows lower flux in the BnA array epoch (Epoch 14), but this is still consistent with 
short-term intrinsic flux variations.
Because the $uv$ distance ranges of the B-array and the BnA-array observations are not very different 
as compared with the C-array observations (Table \ref{tab_obs}), we do not think the drop of the IRS7E flux 
in Epoch 14 is due to the $uv$ sampling. 
After implementing the $\sqrt{u^2 + \nu^2} > $4.4 k$\lambda$ cut (Section \ref{chap_obs}), the source IRS6 
is consistent with a point source. 
As can be seen in Figure \ref{fig_flux1}, the radio variability of IRS6 is dominated by occasional 
short-duration flares, so contamination from the diffuse emission should be negligible.
The rest of the sources in our field are point sources, so flux variability 
can be measured without any bias from the JVLA array configuration.

We found that the fluxes of the four bright sources IRS7W, IRS7E, IRS5, and IRS2 dropped 
by $\sim$10\% simultaneously in Epoch 6 (Table \ref{tab_obs}; Figure \ref{fig_flux2}). 
Because of the larger API rms (Table \ref{tab_obs}) and larger errors of the Gaussian fits, 
we think that this may be due to the loss of coherence caused by the phase noise. 
To some extent the larger error bars can take care of this systematic effect. 
We do not manually correct the fluxes 
because we cannot rule out that this is a real, simultaneous flux drop. 
Nevertheless, we also found that manually correcting the fluxes by 10\% does not qualitatively affect 
our statistical analysis (Section \ref{chap_discussion}).
We did not identify the same issue in other epochs of observations. 
The 1996 fluxes of all sources appear to be systematically lower. 
We hypothesize that it is due to both the poor signal to noise ratio and the loss of phase coherence, 
but it is not completely clear.
Although we will exclude these data points from the following discussion and statistical analysis, 
we found that including them does not change our results qualitatively.

\begin{figure}
\hspace{-0.5cm}
\begin{tabular}{p{4.3cm} p{4.3cm}}
\includegraphics[width=4.8cm]{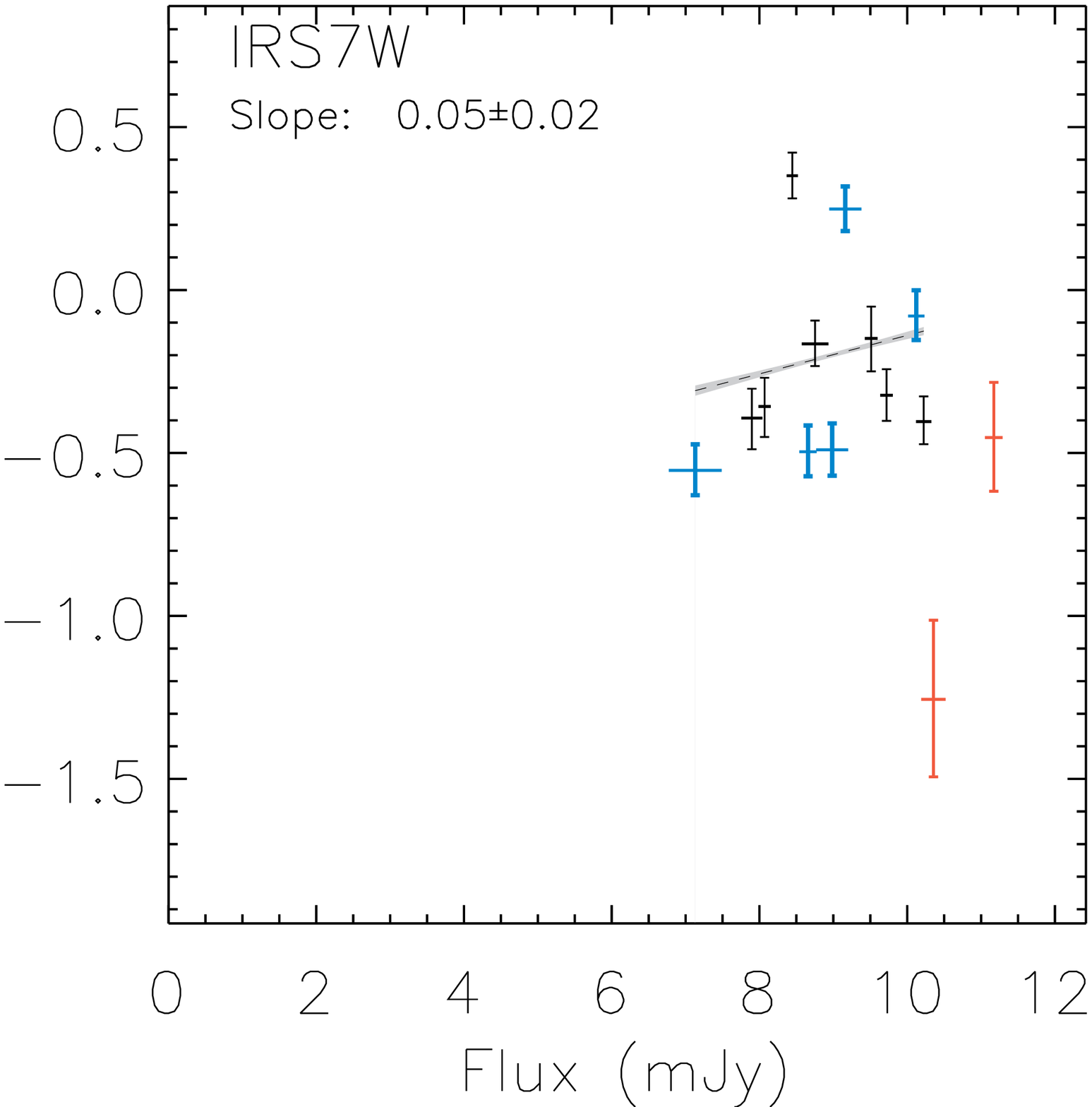} 	&	\includegraphics[width=4.8cm]{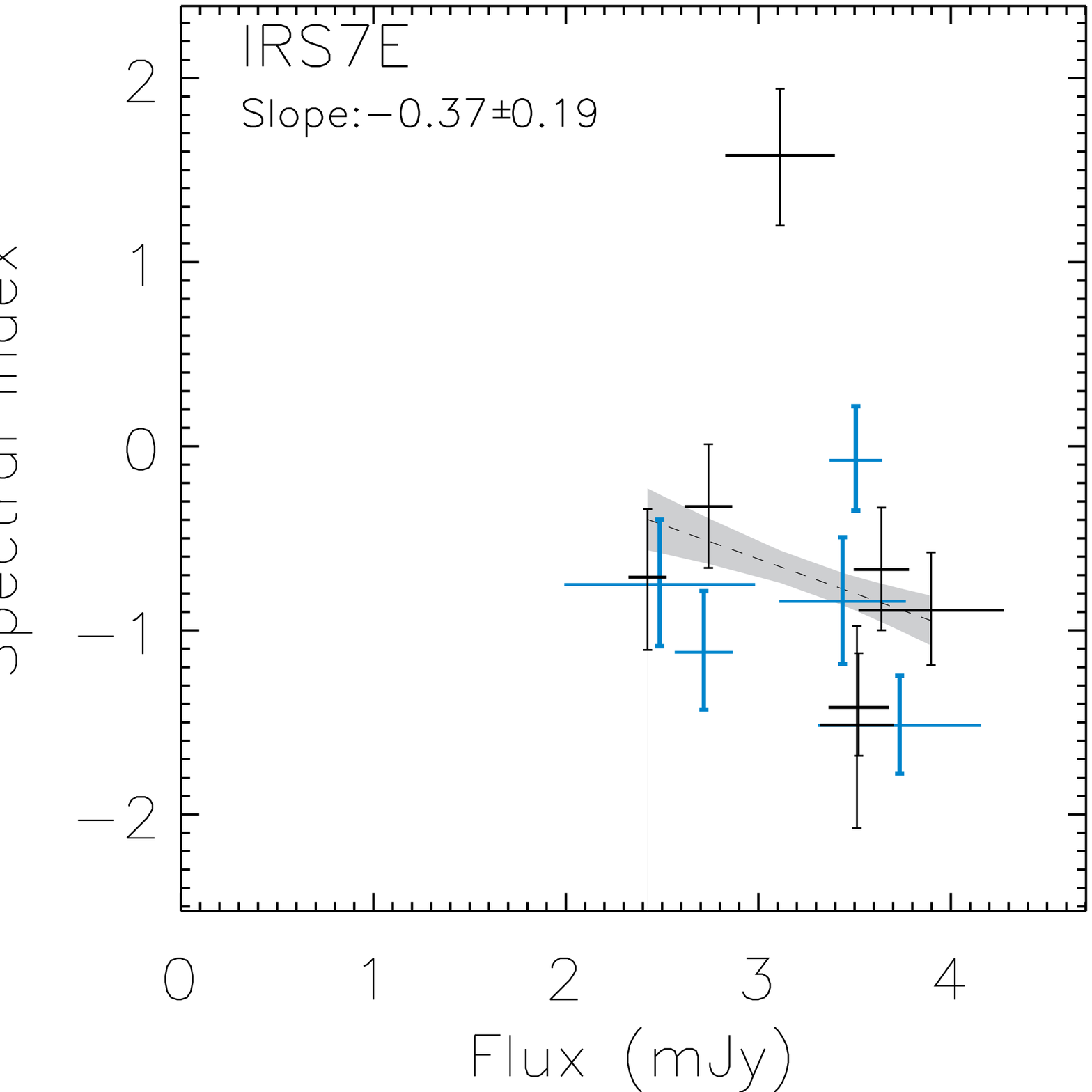}  \\
\end{tabular}

\vspace{-0.6cm}

\hspace{-0.5cm}
\begin{tabular}{p{4.3cm} p{4.3cm}}
\includegraphics[width=4.8cm]{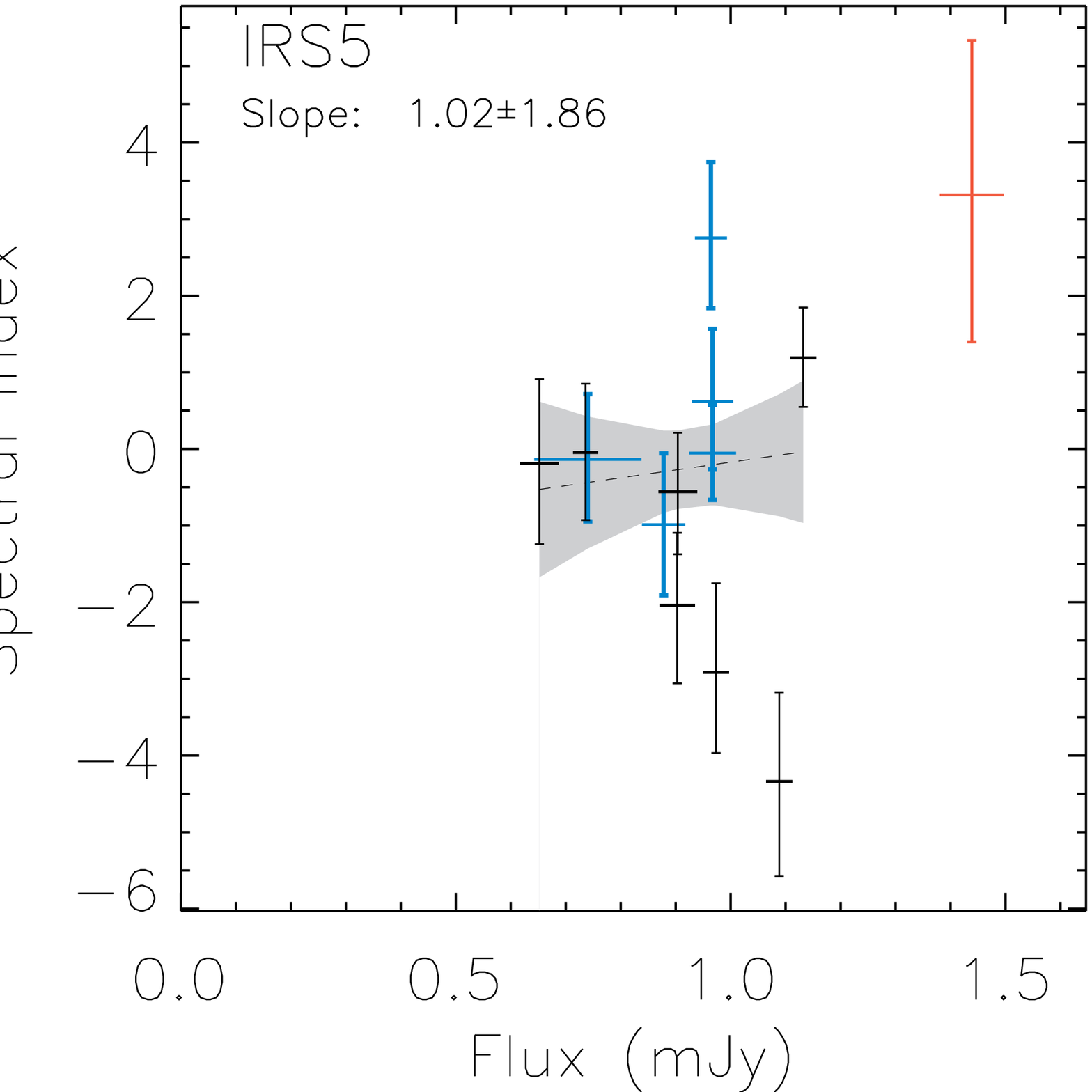}    &  \includegraphics[width=4.8cm]{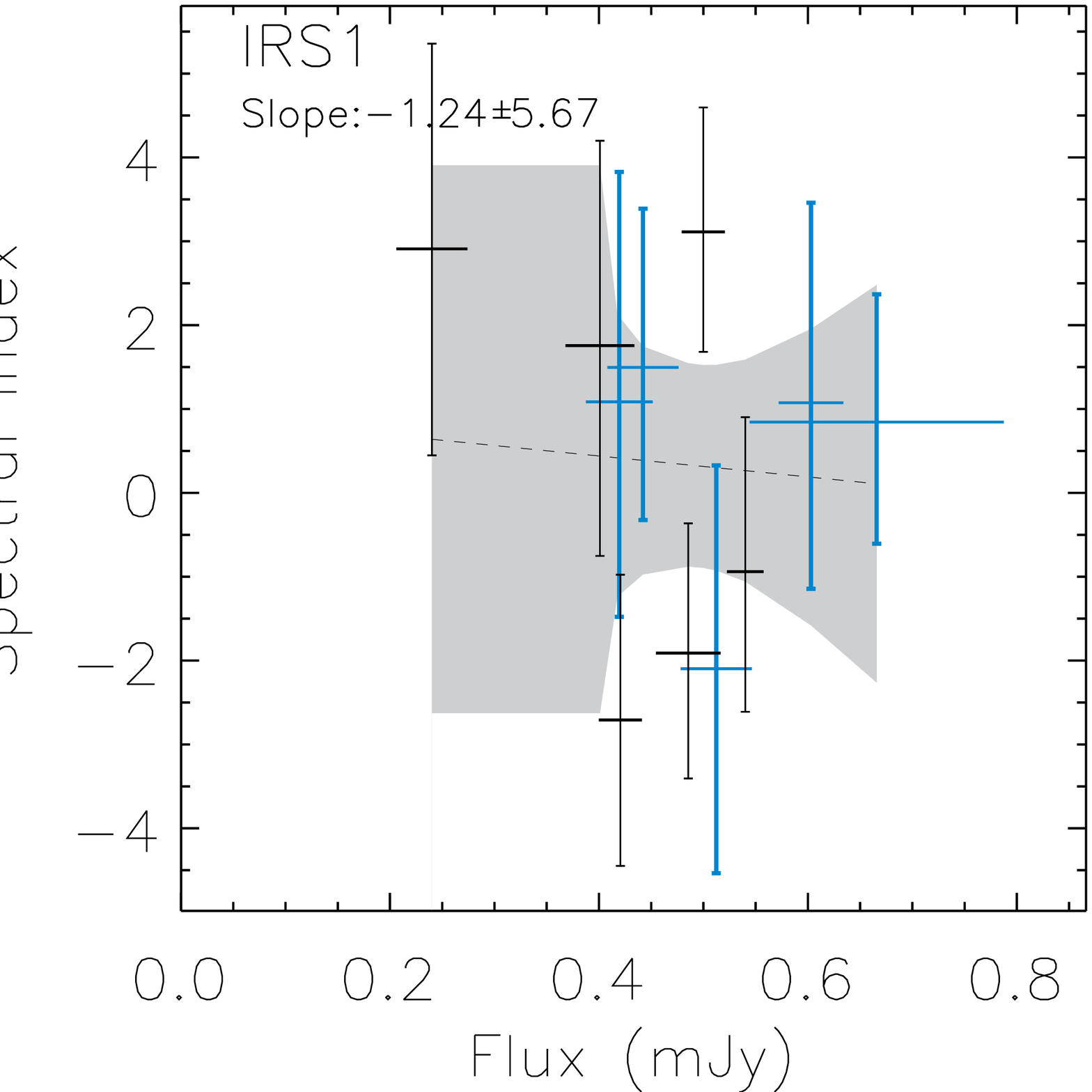}\\
\end{tabular}

\vspace{-0.6cm}

\hspace{-0.5cm}
\begin{tabular}{p{4.3cm} p{4.3cm}}
\includegraphics[width=4.8cm]{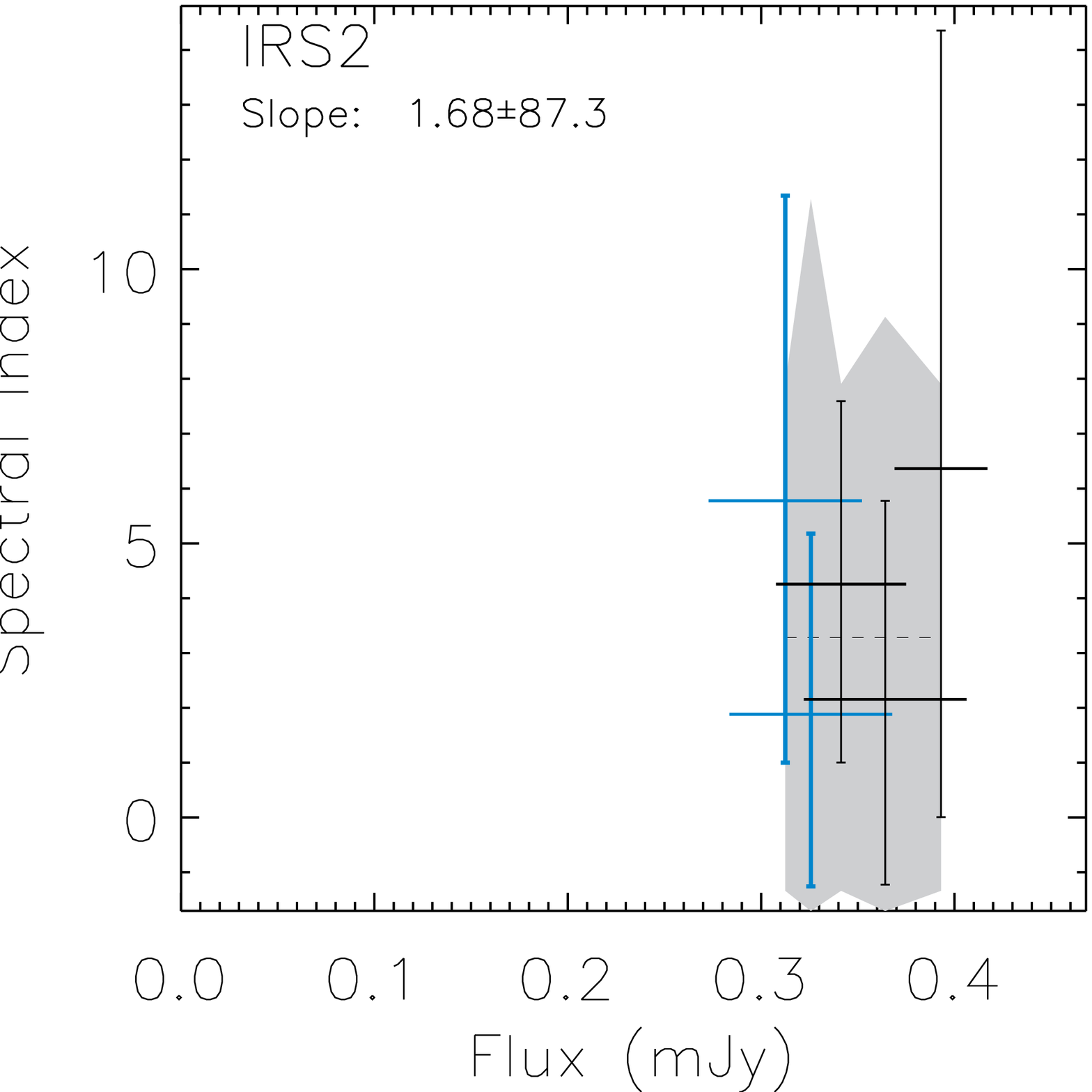}   &  \\
\end{tabular}
\vspace{-0.5cm}
\caption{\footnotesize{
The 3.5-cm spectral indices of the five most significantly detected YSO sources.
The horizontal axis shows the fluxes of the 3.5-cm emission measured from the IF1 data (Table \ref{tab_corr}).
The error bars represent the 1$\sigma$ uncertainties.
The spectral indices derived from the 5 epochs of observations on 2012 March 17 (Table \ref{tab_obs}) are presented in cyan color. 
The spectral indices derived from our JVLA B array and BnA array observations are presented in red color.
The spectral indices derived from other of the 2012 JVLA C array observations are presented in black color.
The dotted line and the shaded area show the results of linear regression for spectral indices derived from 
the JVLA C array observations and the 1$\sigma$ uncertainties returned by the IDL fitting program POLY\_FIT. 
}
}
\label{fig_index}
\end{figure}

Figure \ref{fig_flux1} shows that the 3.5-cm fluxes of the two youngest sources IRS7E and 
IRS7W (Figure \ref{fig_spitzer}) have increased since as early as 1997 January, and 
are now fluctuating around their 2005 February values (Choi et al. 2008).
The mean radio fluxes of these two sources are $\sim$1.5 times higher than the measurements in 1997-1998.
The excess fluxes over the past 14 years are much larger than the measurement errors 
(Figure \ref{fig_flux1}), 
the $<$1-month flux fluctuations (Section \ref{sub_structure}), and all the aforementioned systematic effects.
However, we do not know if any flux variation occurred in these two sources between 2005 and 2012.
The $\lesssim$0.1 mJy sources JVLA2, JVLA3, JVLA4, and IRS5N cannot be detected in individual epochs 
because they are faint (Table \ref{tab_sources}), and also some of them are subject
to a rather large primary beam attenuation (Table \ref{tab_PB}).
In particular, the source JVLA3 resides near the two brightest sources, thus cannot be properly imaged 
given the poor $uv$ coverage in each snapshot JVLA epoch.
The 2012 fluxes of IRS5 fluctuated about its 2005 value (Choi et al. 2008), 
which is $\sim$1/3 of the maximum of 3.3 mJy detected in 1998 (Forbrich et al. 2006).
The 2012 fluxes of two of the three sources in between the Class I and Class II stages (Figure \ref{fig_spitzer}, 
IRS1 and IRS2) are pretty consistent with all earlier observations.
They show variability on timescales of 10$^{-2}$ to 10$^{2}$ days.
The source IRS6 is only detected in occasional flares, similar to the Class II and the Class III sources 
T\,CrA, R\,CrA, and JVLA1.
The source R\,CrA (the brightest in the near infrared)
occasionally falls below our detection limits ($<$2$\sigma$) in our 2012 observations,  
while it was detected in all epochs of earlier observations.
However, from Figure \ref{fig_flux2}, the low state of R\,CrA may only last for $<$30 minutes. 
Because all previous observations required at least several hours of on source integration to achieve the 
adequate signal-to-noise (S/N) ratio, they might not be sensitive to the low states of R\,CrA. 

The 2012 fluxes of the extragalactic source B5 fluctuated within the same range as in the 
earlier observations for timescales $>$0.1 day.
The very small differences ($\pm$6\%) in the measured flux of B5 among the 5 epochs of observations on March 17, 
and between the 2 epochs of observations on March 22 (Figure \ref{fig_flux2}, Table \ref{tab_obs}) may be 
explained by the characteristic $>$10$^{4}$ s variability of the innermost accretion flow around a 
supermassive black hole (e.g. Miniutti et al. 2006), which in fact makes the gain calibration of JVLA data 
possible 
(Section \ref{chap_obs}).
The small flux variations observed in B5 on $<$0.1-day timescales may provide an upper 
limit on the flux measurement uncertainties.

\subsection{The Stokes \textit{I} 8.2-9.1 GHz Spectral Index}
\label{sub_index}
We provide a preliminary analysis of the spectral index and the fluctuations of the spectral index of bright sources by comparing the radio fluxes measured from spectral window 1 and spectral window 8 (Table \ref{tab_corr}).
These two spectral windows have low noise levels, and are adequately separated in frequency. 
Ideally, the spectral index analysis should incorporate the fluxes from higher frequency spectral windows.  
This is presently hindered by the higher noise in those spectral windows, and our inability to correctly combine the data (Section \ref{chap_obs}). 

\begin{table}\scriptsize{
\vspace{0cm}
\caption{\footnotesize{The earlier measurements of spectral indices between 6 cm and 3 cm.}}
\label{tab_preindex}
\vspace{0.3cm}
\hspace{1.2cm}
\begin{tabular}{lcc}\hline\hline
Source name		&			1996 Dec 29$^{\mbox{\tiny{a}}}$		&		1998 Jan 9/10$^{\mbox{\tiny{b}}}$		\\
\hline
IRS7A				&				0.04$\pm$0.05						&		0.19$\pm$0.04								\\
B9						&				-0.36$\pm$0.08								&		$\cdots$									\\
FPM13				&				-0.2$\pm$0.4					&		$\cdots$											\\
IRS7B				&					-1.41$\pm$0.09					&	0.38$\pm$0.02									\\
IRS5					&					-0.4$\pm$0.4					&		-0.12$\pm$0.07									\\
IRS1					&								0.9$\pm$0.3						&			$\cdots$							\\
IRS2					&			$\cdots$										&		$\cdots$											\\
\hline
\end{tabular}
}

\vspace{0.1cm}
\footnotesize{Note.--- 
}
\par
\scriptsize{
\begin{itemize}
\item[$^{\mbox{\scriptsize{a}}}$] Derived from the VLA observations reported in Choi et al. (2008). \vspace{-0.15cm}
\item[$^{\mbox{\scriptsize{b}}}$] Derived from the ATCA observations reported in Miettinen et al. (2008). 
We note that these ATCA observations cannot resolve FPM15 from IRS7B; and cannot resolve FPM10 and FPM13 from 
IRS7A and B9. 
\end{itemize}
\vspace{-0.1cm}}
\end{table}

We smoothed the image of spectral window 8 to the same angular resolution of spectral window 1 before measuring 
the fluxes. 
The images of spectral windows 1 and 8 are subject to a higher noise than the broad band images 
(Section \ref{sub_Ivar}, Table \ref{tab_obs}) because of the smaller bandwidth.
Therefore, we trimmed both images to the 4$\sigma$ level to avoid confusion by noise. 
In some epochs, the sources cannot be detected in the images of spectral window 8 after trimming, thus the spectral indices were not derived. 
The obtained spectral indices, if available, are presented with the 3.5-cm fluxes in Figure \ref{fig_index}.
Because the spectral window 1 is more sensitive and is subject to a smaller primary beam attenuation, 
the measurements presented in Figure \ref{fig_index} preferentially picked positive 
spectral indices for weak ($<$0.5 mJy) sources. 
We performed linear regressions to the results presented in Figure \ref{fig_index}, but only 
for those derived from the C array observations (Table \ref{tab_obs}). 
This is because the effects of the 4$\sigma$ trimming can be different for images with  
different angular resolution. 

For comparison, in Table \ref{tab_preindex} we quote the spectral indices
previously measured  between 6 cm and 3 cm.
The quoted spectral indices were derived from observations taken on the same date, but
not exactly simultaneous.
Although IRS2 is too faint to obtain a meaningful constraint on its spectral index, 
the rest of the measurements presented in Figure \ref{fig_index} are consistent with 
earlier observations.
In particular, our measurements of the spectral index of IRS7E 
($\alpha^{\mbox{\scriptsize{obs}}}_{\mbox{\tiny{IRS7E}}}$) vary within a range consistent with the 
previously reported spectral index of IRS7B.
$\alpha^{\mbox{\scriptsize{obs}}}_{\mbox{\tiny{IRS7E}}}$ shows a general trend of 
being more negative when the 3.5 cm flux is higher, but it has an exception 
with $\alpha^{\mbox{\scriptsize{obs}}}_{\mbox{\tiny{IRS7E}}}$$\sim$$+$2.
$\alpha^{\mbox{\scriptsize{obs}}}_{\mbox{\tiny{IRS7E}}}$ converged towards $\sim$0 during its lowest flux status.
It is not straightforward to understand the variations of the spectral index of IRS7W because of its multiplicity.
From Figure \ref{fig_index}, we see that for the majority time, 
$\alpha^{\mbox{\scriptsize{obs}}}_{\mbox{\tiny{IRS7W}}}$ stayed within the range $\pm$0.5, and became less 
than $-$1 in one event of flux increase. 
For IRS5, similarly to IRS7E, $\alpha^{\mbox{\scriptsize{obs}}}_{\mbox{\tiny{IRS5}}}$ 
also converged towards $\sim$0 during its lowest flux status.
The distribution of $\alpha^{\mbox{\scriptsize{obs}}}_{\mbox{\tiny{IRS5}}}$, however, 
cannot be represented by a linear relation, but is rather bimodal.
For IRS1, $\alpha^{\mbox{\scriptsize{obs}}}_{\mbox{\tiny{IRS1}}}$ fluctuates around 0, 
with a marginal trend to be more negative when the flux is higher.
More discussion about the spectral indices is deferred to Section \ref{sub_interpretation}.

\subsection{The Stokes \textit{V} Flares}
\label{sub_stokesV}

\begin{figure}
\begin{tabular}{p{9cm} }
	\\
\end{tabular}

\vspace{-0.8cm}

\hspace{-1.7cm}
\begin{tabular}{p{9cm} }
\includegraphics[width=11.5cm]{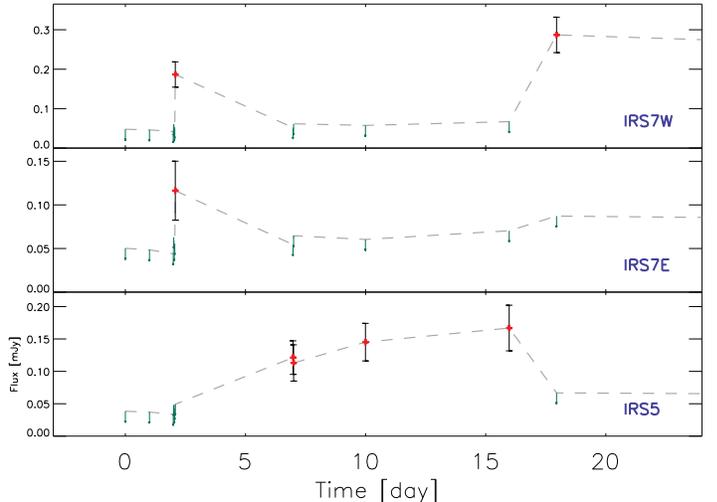} 	\\
\end{tabular}
\vspace{-0.5cm}
\caption{\footnotesize{
The detected 3.5 cm Stokes \textit{V} emission (see also Table \ref{tab_stokesV}) in the 2012 observations (Table \ref{tab_obs}). 
For particular 2012 epochs when the sources cannot be detected, the 2$\sigma$ upper limit is given in green downward arrows. 
Otherwise we show the $\pm$1 $\sigma$ uncertainties in black error bars.
}
}
\label{fig_stokesV}
\end{figure}

We used the method introduced in Section \ref{sub_Ivar} to measure the fluxes of the 3.5-cm Stokes 
\textit{V} emission. 
The results are shown in Figure \ref{fig_stokesV} and Table \ref{tab_stokesV}.
We detected Stokes \textit{V} flares towards the three Class 0/I sources IRS7E, IRS7W, and IRS5.
From the position of the detection, the Stokes \textit{V} flares observed in IRS7W are likely to be 
dominated by the component IRS7A, rather than by B9 or any component. 
However, given the angular resolution of our observations (Table \ref{tab_obs}), 
we cannot rule out that B9 contributed partially.  

Stokes \textit{V} emission from IRS7A and IRS5 was also reported in previous VLA observations (Feigelson et al. 1998; Forbrich et al. 2006; Choi et al. 2008, 2009).
While the Stokes \textit{V} flares of IRS5 were observed to have durations $>$30 days 
(Forbrich et al. 2006; Choi et al. 2008, 2009), the Stokes \textit{V} emission from IRS7A was 
detected in only one previous observation (Choi et al. 2008). 
Our results are qualitatively similar to previous reports, 
and will be briefly discussed in Section \ref{sub_interpretation}.

\begin{table}\scriptsize{
\vspace{-0.1cm}
\caption{\footnotesize{The 2012 detections of Stokes \textit{V} emission and the polarization percentage.}}
\label{tab_stokesV}
\vspace{0.0cm}
\hspace{0.0cm}
\begin{tabular}{c | cc | cc | cc}\hline\hline
				&		\multicolumn{2}{ c| }{IRS7A}		&		\multicolumn{2}{ c| }{IRS7B}		&		\multicolumn{2}{ c }{IRS5}  	\\
				&	V			&		V/I	$^{\mbox{\tiny{a}}}$						&			V			&		V/I	$^{\mbox{\tiny{a}}}$				&		V			&		V/I					\\
Epoch		&	($\mu$Jy)		&		(\%)						&		($\mu$Jy)			&		(\%)				&		($\mu$Jy)		&		(\%)				\\\hline
7				&	183$\pm$32	&		2.2$\pm$0.4	&	113$\pm$33	&		4.5$\pm$1.3	&			$\cdots$			&		$\cdots$				\\
8				&	$\cdots$			&		$\cdots$			&	$\cdots$			&		$\cdots$			&			121$\pm$26	&		10.7$\pm$2.3		\\
9				&	$\cdots$			&		$\cdots$			&	$\cdots$			&		$\cdots$			&			113$\pm$28	&		10.4$\pm$2.6		\\
10				&	$\cdots$			&		$\cdots$			&	$\cdots$			&		$\cdots$			&			144$\pm$29	&		19.7$\pm$3.9		\\
11				&	$\cdots$			&		$\cdots$			&	$\cdots$			&		$\cdots$			&			161$\pm$35	&		19.5$\pm$4.1		\\
12				&	281$\pm$44	&		3.4$\pm$0.5	&	$\cdots$			&		$\cdots$			&			$\cdots$			&		$\cdots$				\\
\hline
\end{tabular}
}

\vspace{0.1cm}
\hspace{0.5cm}
\footnotesize{Note.---  These measurements incorporate all data in IF1 (Table \ref{tab_corr}). 
These data are plotted in Figure \ref{fig_stokesV}.
}
\par \vspace{-0.2cm}
\scriptsize{
\begin{itemize}
\item[$^{\mbox{\scriptsize{a}}}$] Lower limits to total polarization percentage. 
\end{itemize}
\vspace{0.1cm}}
\end{table}


\section{Discussion}
\label{chap_discussion}
We examine the statistics of the measured Stokes \textit{I} fluxes in the 2012 JVLA observations in Section \ref{sub_biweight}.
We compare our observations with the earlier VLA observations in Section \ref{sub_structure}.
Our tentative interpretation of the observational results is provided in Section \ref{sub_interpretation}.

\subsection{Statistics of Stokes \textit{I} Emission in 2012}
\label{sub_biweight}

\begin{figure*}
\begin{tabular}{p{9cm} }
	\\
\end{tabular}

\vspace{-0.4cm}

\hspace{-0.6cm}
\begin{tabular}{p{6cm} p{6cm} p{6cm} }
\includegraphics[width=7cm]{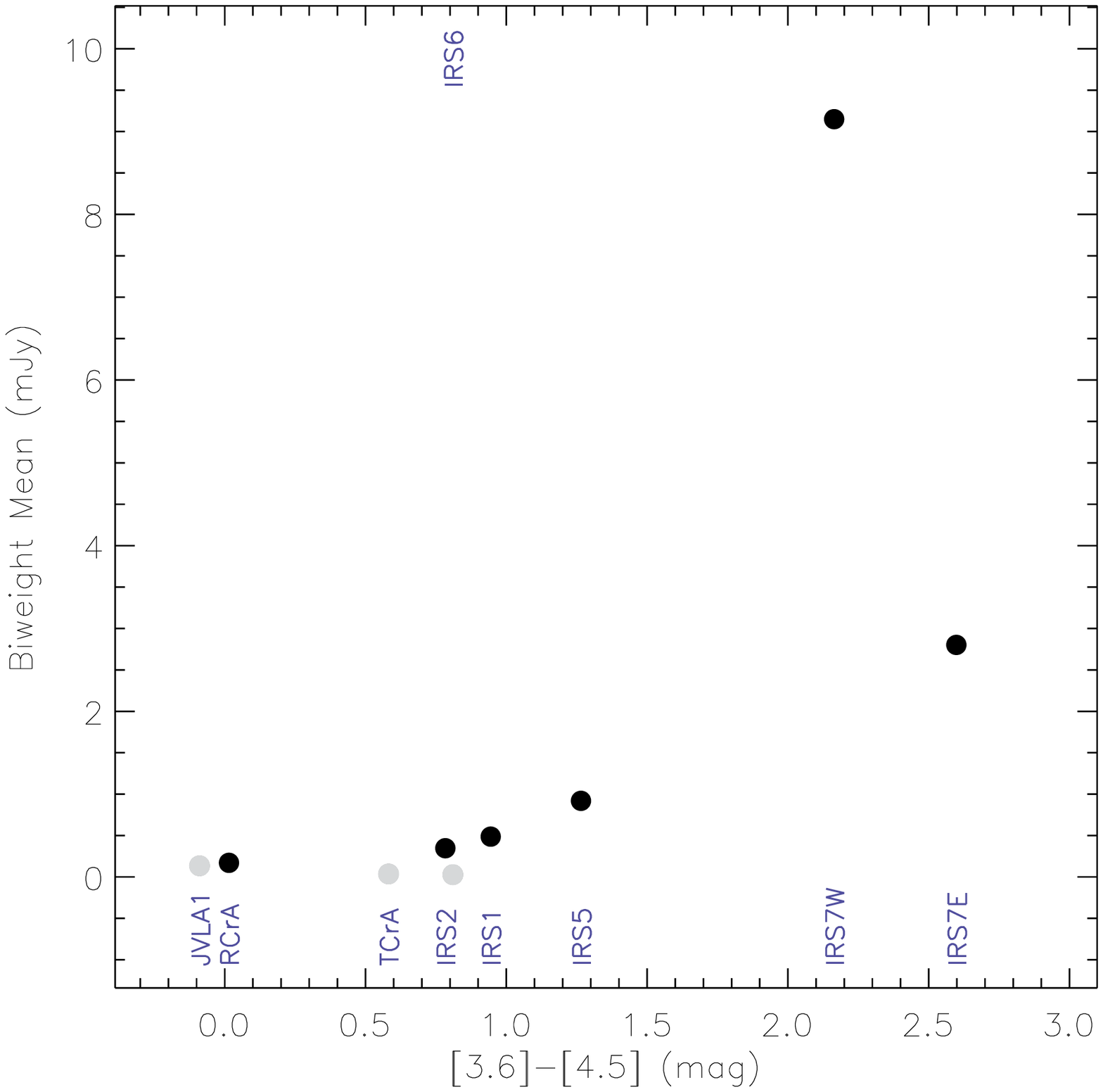}		&   	\includegraphics[width=7cm]{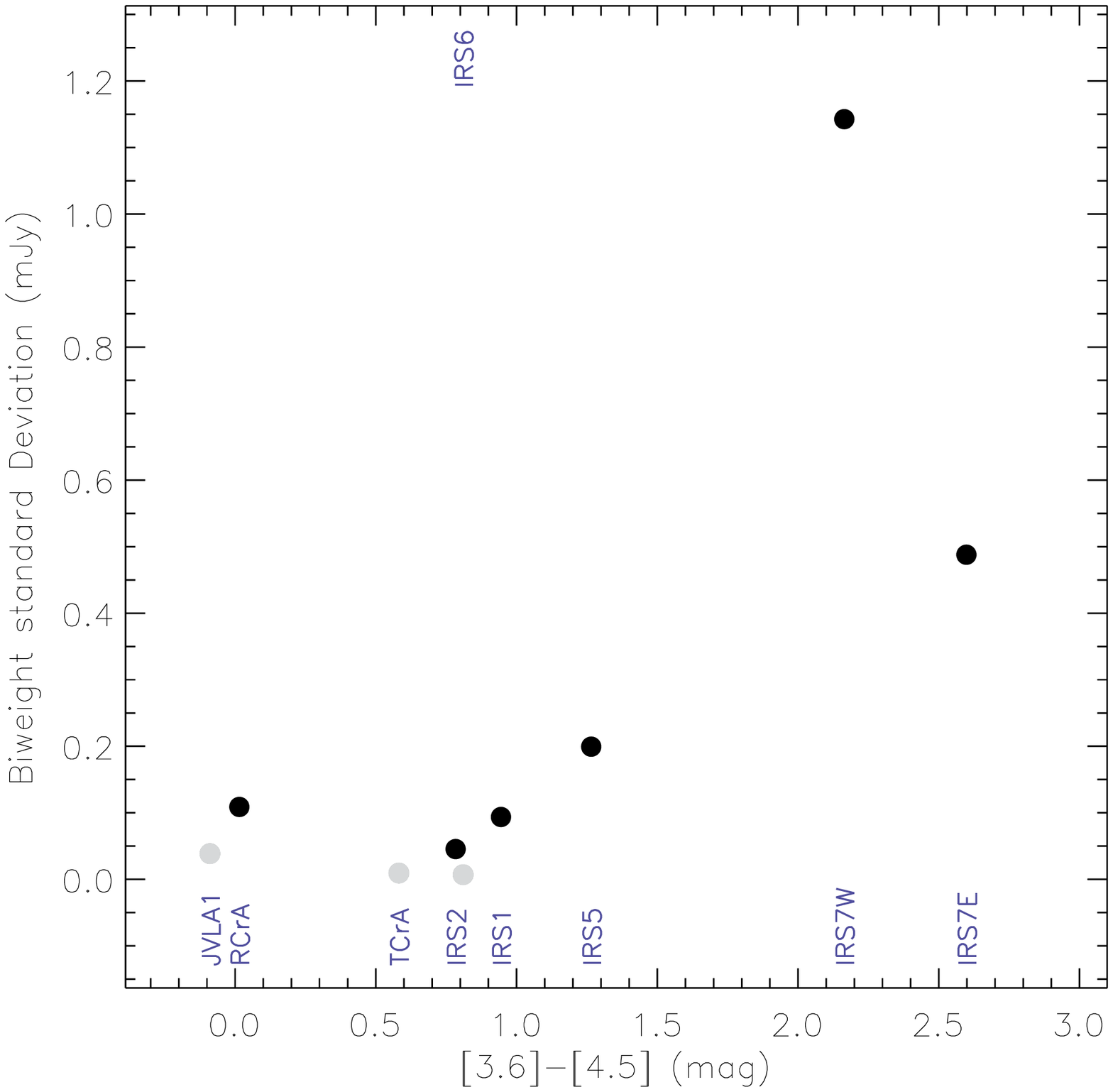}	&  \includegraphics[width=7cm]{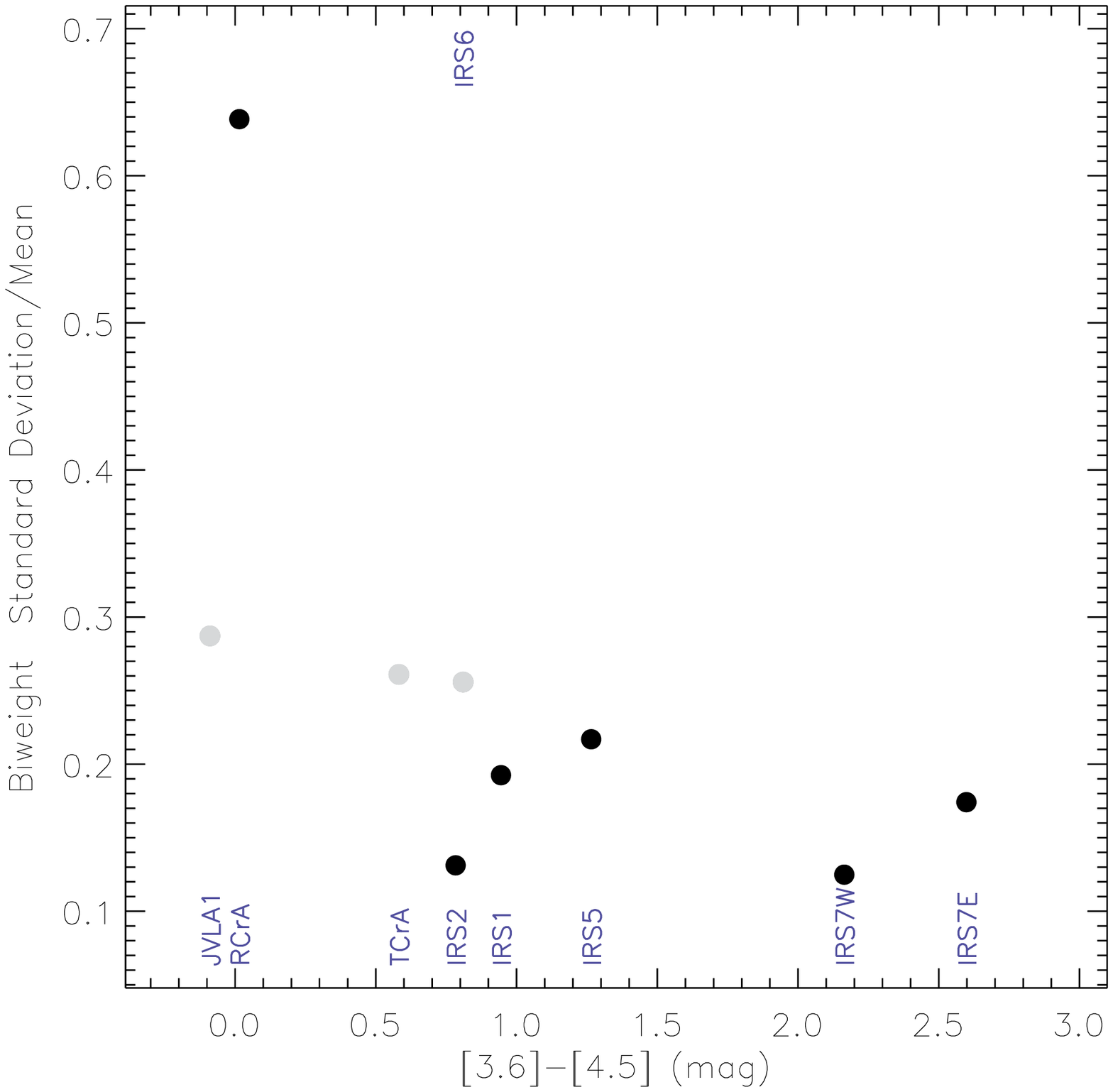}\\
\end{tabular}

\vspace{-1cm}
\hspace{1.5cm}
\begin{tabular}{p{6cm} p{6cm} p{6cm} }
\includegraphics[width=7cm]{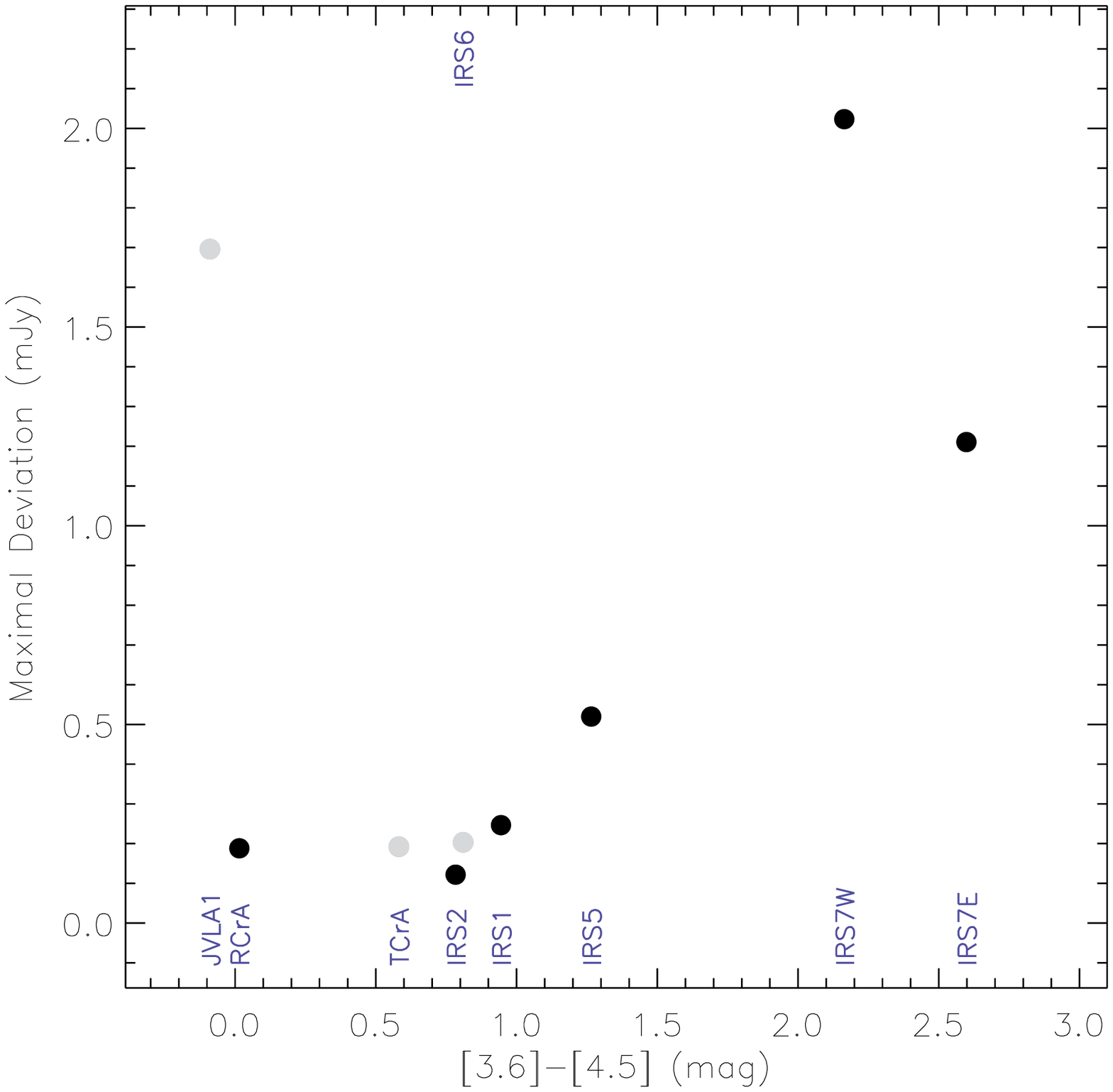}		&   	\includegraphics[width=7cm]{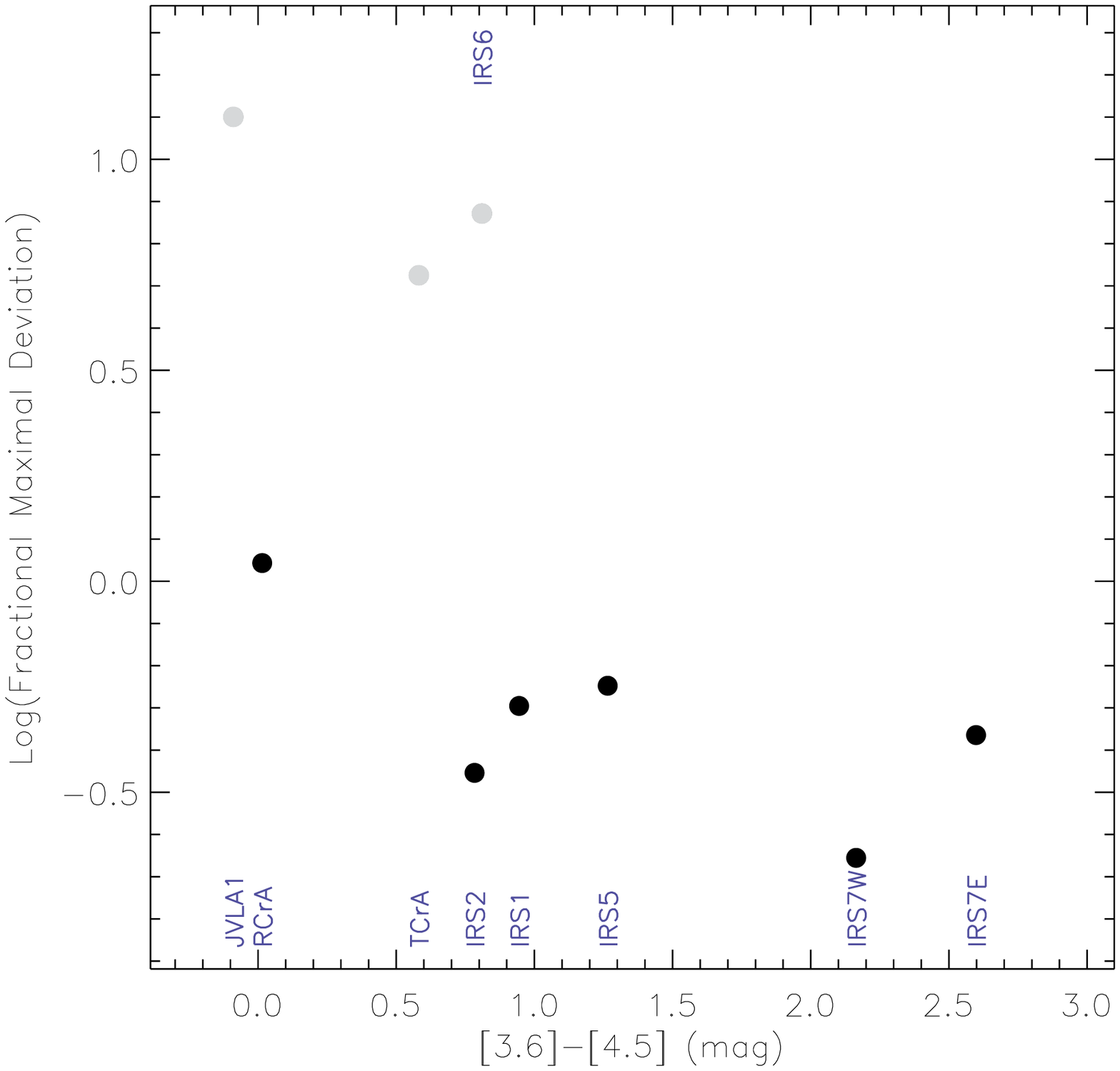}\\

\end{tabular}
\vspace{-0.2cm}
\caption{\footnotesize{
The biweight mean of the 3.5-cm Stokes \textit{I} flux $F$, the (biweighted) standard deviation 
($\sigma_{\mbox{\scriptsize{bw}}}$) of the flux, the fractional standard deviation
$\sigma_{\mbox{\scriptsize{bw}}}$/$F$, the absolute maximal flux deviation relative to the biweighted mean 
$\Delta F^{\mbox{\scriptsize{max}}}$, and the fractional maximal flux deviation 
$\Delta F^{\mbox{\scriptsize{max}}}$/F. 
The statistics in these diagrams incorporate all the 2012 IF1 data summarized in Tables \ref{tab_obs} and 
\ref{tab_corr}.
The horizontal axis can be used as an indicator of the YSO evolutionary stage 
(see also Figure \ref{fig_spitzer}).
We note that the three sources TCrA, IRS6, and JVLA1 were only occasionally detected (presented in gray dots). 
The values of $F$ and $\sigma_{\mbox{\scriptsize{bw}}}$ for these sources are therefore upper limits 
($\sigma_{\mbox{\scriptsize{bw}}}$/$F$ is not very meaningful), 
and the values of $\Delta F^{\mbox{\scriptsize{max}}}$/F are lower limits. 
$\sigma_{\mbox{\scriptsize{bw}}}$ can represent the uncertainty in the biweighted mean as well as its trend.
}
}
\label{fig_statistics1}
\end{figure*}

\begin{figure*}
\begin{tabular}{p{9cm} }
	\\
\end{tabular}

\vspace{-0.8cm}

\hspace{-0.6cm}
\begin{tabular}{p{6cm} p{6cm} p{6cm} }
\includegraphics[width=7cm]{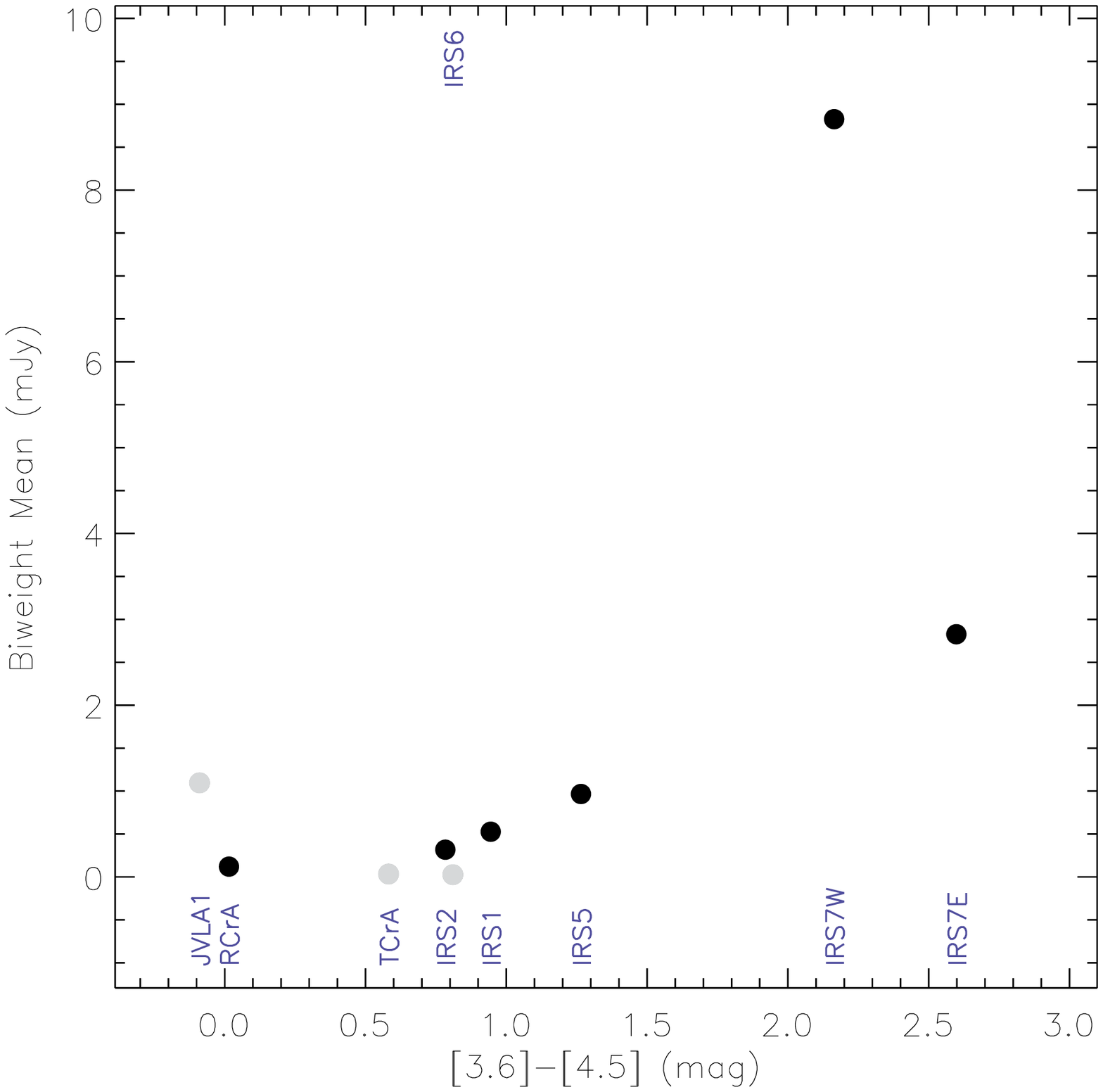}		&   	\includegraphics[width=7cm]{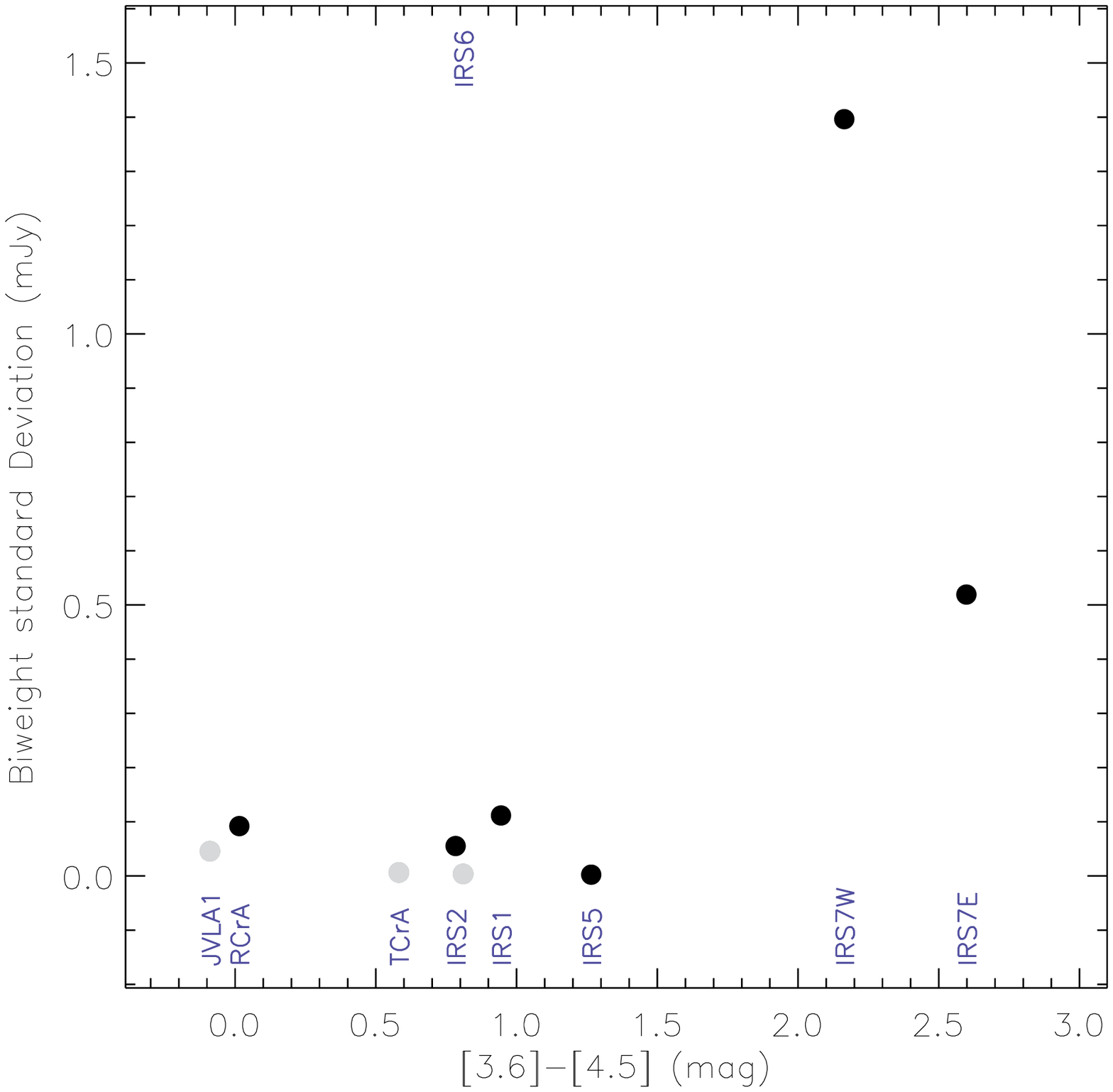}	&  \includegraphics[width=7cm]{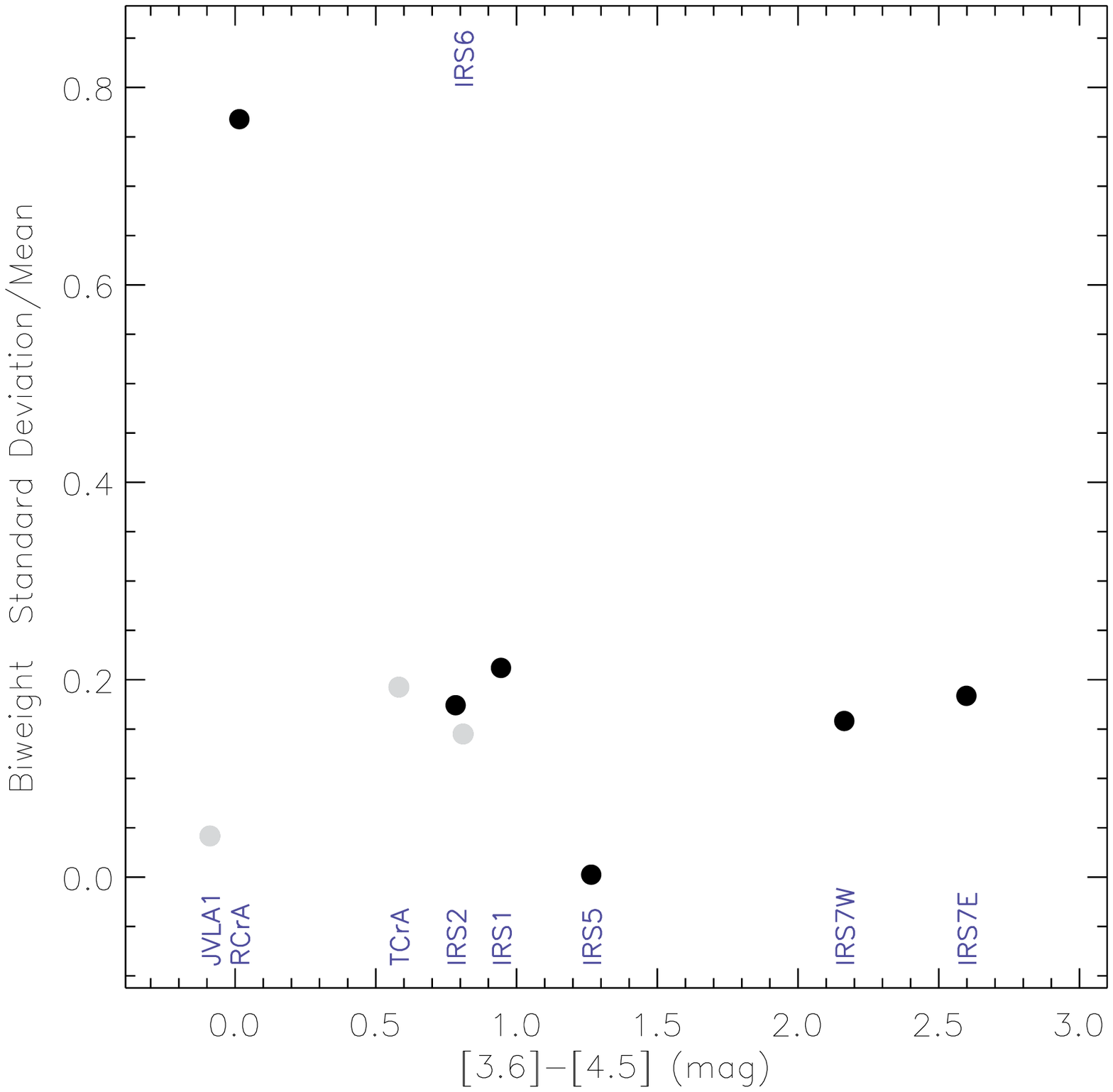}\\
\end{tabular}

\vspace{-1cm}
\hspace{1.5cm}
\begin{tabular}{p{6cm} p{6cm} p{6cm} }
\includegraphics[width=7cm]{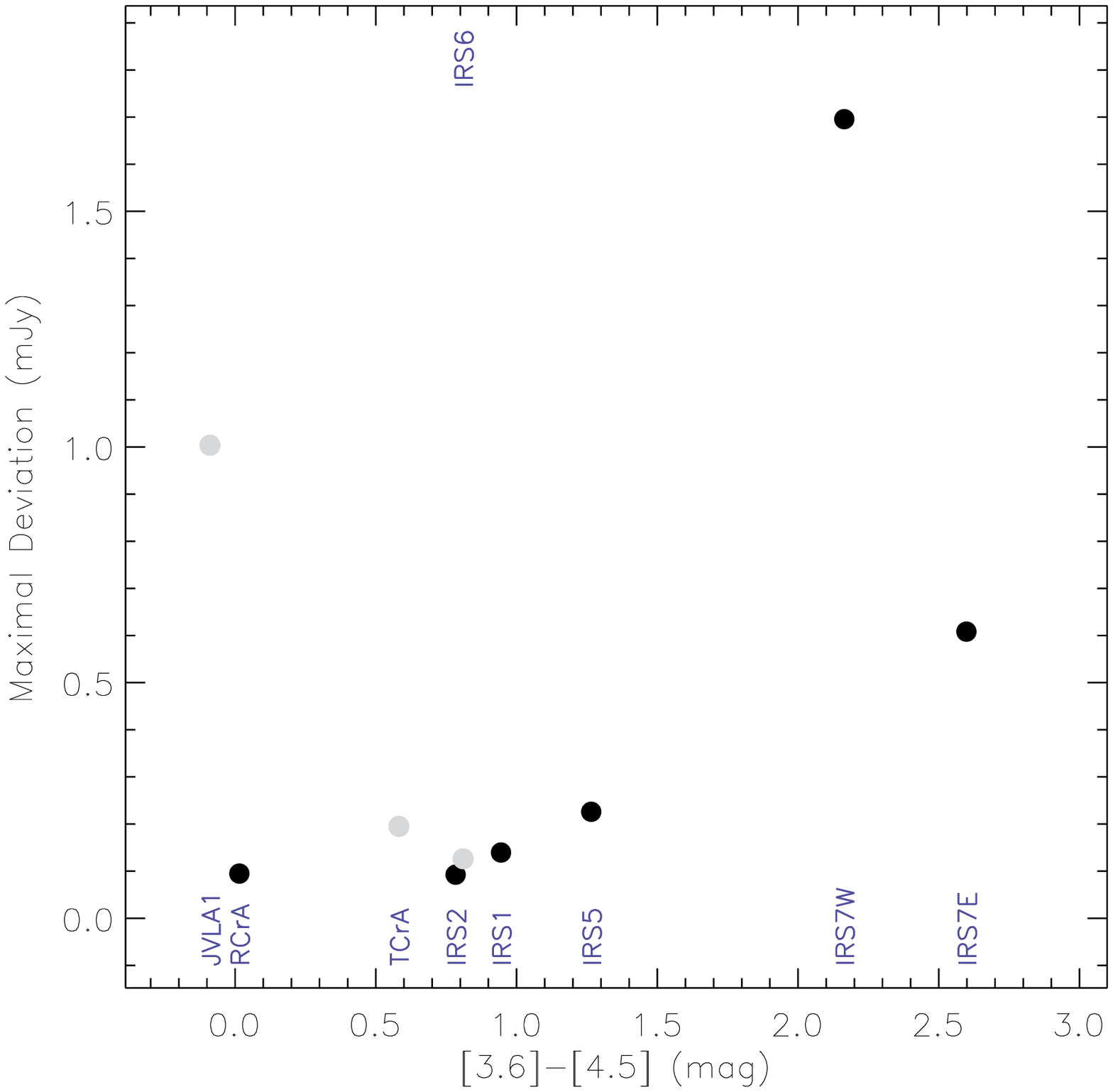}		&   	\includegraphics[width=7cm]{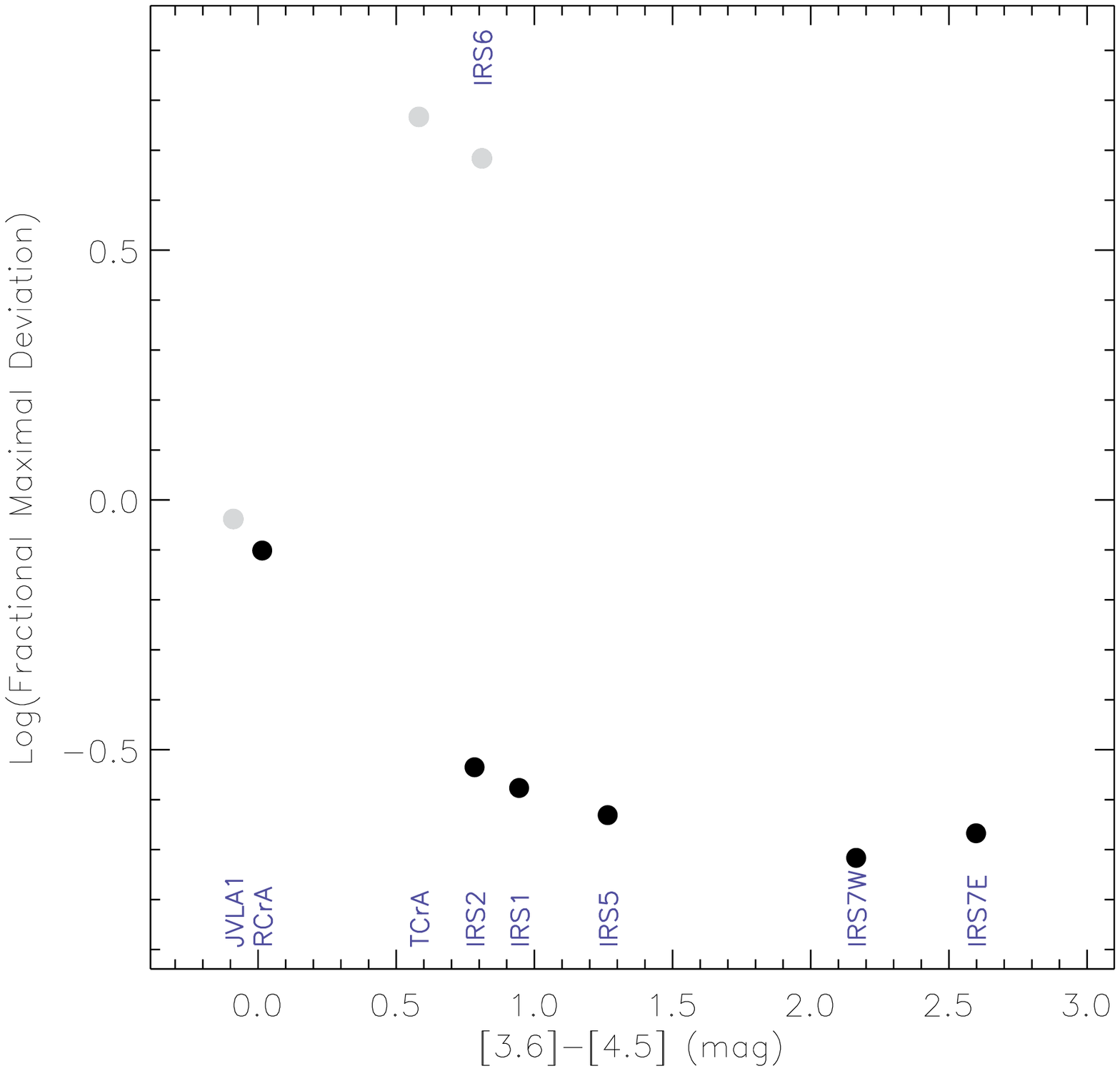}\\

\end{tabular}
\vspace{-0.2cm}
\caption{\footnotesize{
Similar to Figure \ref{fig_statistics1}, however, we only incorporate the 5 epochs 
taken on 2012 March 17 in the analysis. 
The two sources TCrA and IRS6 were only detected once in these epochs. 
The values of $F$ and $\sigma_{\mbox{\scriptsize{bw}}}$ for these sources are therefore upper limits, 
and the values of $\Delta F^{\mbox{\scriptsize{max}}}$/F are lower limits. 
The biweight mean and standard deviation of JVLA1 in this figure are significant, because it is detected in
more than half of the epochs in these observations. 
}
}
\label{fig_statistics2}
\end{figure*}

We used the \textit{biweight} method in robust statistics (Hoaglin et al. 1983) to estimate the steady flux 
levels and the dispersions of the fluxes.
This method is advantageous because it can objectively lower the weights or reject the measurements 
which are largely deviated from the mean value. 
Therefore, the steady flux levels and dispersions derived using this method are less biased by  
fast flaring or fading of the YSOs as well as the occasional impact of 
potential calibration issues (e.g. see Section \ref{sub_Ivar}). 
We used the \texttt{BIWEIGHT\_MEAN} routine in the IDL Astronomy User's Library (Landsman 1993) to iteratively 
estimate the biweight mean ($F$), the biweighted standard deviation ($\sigma_{\mbox{\scriptsize{bw}}}$), 
and the normalized biweighted standard deviation $\sigma_{\mbox{\scriptsize{bw}}}$/$F$ for the 3.5-cm Stokes  
\textit{I} fluxes (Section \ref{sub_Ivar}).
For the non-detections in individual epochs, we use the 1$\sigma$ rms noise in one synthesized beam 
(i.e., units in mJy\,beam$^{-1}$). 
The values derived from the JVLA observations taken in 2012 from March 15 to September 12 (Table \ref{tab_obs}), and derived from the 5 epochs of JVLA observations taken on 2012 March 17, are given in Figure \ref{fig_statistics1} and \ref{fig_statistics2}, respectively. 
In the same figures, we also provide the (fractional) maximal deviation of the fluxes 
(i.e., $\Delta F^{\mbox{\scriptsize{max}}}$ and $\Delta F^{\mbox{\scriptsize{max}}}$/F) in these two periods. 

The results in Figure \ref{fig_statistics1} and \ref{fig_statistics2} show similar trends, 
suggesting that some of the detected flux variations within $\sim$10 to $10^2$ days 
can be attributed to phenomena with shorter durations.
Observations separated by days may also be capable of characterizing (at least partially) 
the mean flux level and the variability in shorter periods.
In addition, adding or removing a few records does not seem to impact the statistics qualitatively, thanks to the moderate resistance of the biweight method.
We note that the steady flux level of JVLA1 is potentially comparable to that of R\,CrA.
However, JVLA1 was only detected when it flared to $\gtrsim$1 mJy because of its 
large primary beam attenuation (Table \ref{tab_PB}).
The Class II source T\,CrA, and the Class I/II source IRS6 are the least active sources at all 
timescales (Figure \ref{fig_flux1}, \ref{fig_flux2}). For these two sources, 
only the upper limit of the biweight mean 3.5-cm fluxes can be given. 
The faintest sources JVLA2, JVLA3, JVLA4, and IRS5N, 
which cannot be detected in any of the individual epochs, 
are omitted from Figures \ref{fig_statistics1} and \ref{fig_statistics2}, 
but will be discussed in Section \ref{sub_interpretation}.

For the more reddened YSOs, we found that although $\sigma_{\mbox{\scriptsize{bw}}}$ is the largest, 
$F$ is large enough such that $\sigma_{\mbox{\scriptsize{bw}}}$/$F$ is $<0.2$.
The biweight mean of IRS7W appears to be far larger than for the rest of the sources, 
most likely due to the fact that we cannot resolve the multiple embedded YSOs and jet knots 
(Choi et al. 2004), and also because IRS7W is currently in a high state (Figure \ref{fig_flux1}).
The 3.5 cm fluxes of the less reddened YSOs have larger variations compared to their 
steady flux level, as seen from their
larger $\sigma_{\mbox{\scriptsize{bw}}}$/$F$ and $\Delta F^{\mbox{\scriptsize{max}}}$/F.

\begin{figure}
\begin{tabular}{p{9cm} }
	\\
\end{tabular}

\vspace{-1.0cm}

\hspace{0.3cm}
\begin{tabular}{p{9cm} }
\includegraphics[width=8.5cm]{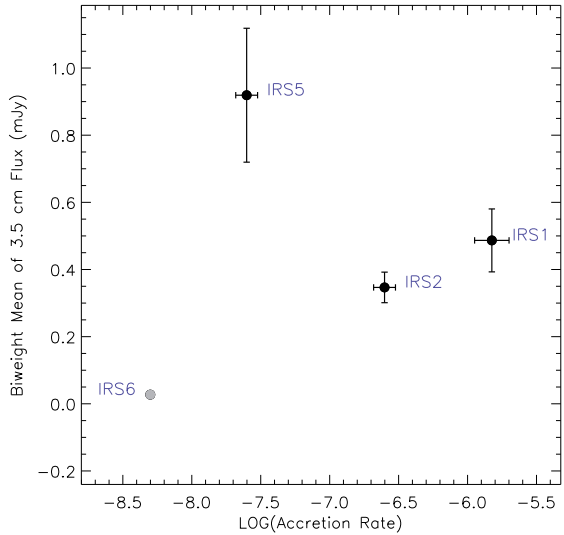} 	\\
\end{tabular}
\vspace{-0.4cm}
\caption{\footnotesize{
The biweight mean of the 3.5-cm Stokes \textit{I} fluxes (mJy) and the VLT-ISAAC measured YSO accretion 
rates ($M_{\odot}$ yr$^{-1}$) presented in Nisini et al. (2005).
The 3.5 cm-flux and the accretion rate of IRS6 are both upper limits. 
Note that the accretion rate and the radio fluxes are not simultaneously observed. 
}
}
\label{fig_accretion}
\vspace{0.2cm}
\end{figure}

The accretion rates of the four sources IRS5, IRS1, IRS2, and IRS6, in  
2002, were constrained by a near infrared spectroscopic survey 
(Nisini et al. 2005)\footnote{Nisini et al. (2005) estimated the differences between the observed 
bolometric luminosity and the stellar luminosity. We refer to this original paper for uncertainties in their 
estimates. We are planning to obtain new values of accretion rates in future programs.}.
We compare the biweight mean of their 3.5-cm Stokes \textit{I} fluxes from March 15 to September 12  
with the reported accretion rates (Figure \ref{fig_accretion}).
A weak point of this comparison is that the radio fluxes from all available observations are separated from the 
observations of the accretion rates by $\sim$3-10 years.
However, based on the monitoring observations of the accretion rates on a large number of Class II YSOs 
(Costigan et al. 2012), we hypothesize that the accretion rate of the YSOs may not change too much 
on this timescale except for the case of accretion instabilities 
(e.g., Findeisen et al. 2013, and references therein).
In the three Class I/II sources (IRS1, IRS2, and IRS6; see Figure \ref{fig_spitzer}), IRS1 was 
observed to have the highest accretion rate ($\dot{M_\mathrm{in}}\sim2\times10^{-6}$ M$_{\odot}$ yr$^{-1}$).
The accretion rate of IRS 2 was $\dot{M_\mathrm{in}}\sim3\times10^{-8}$ M$_{\odot}$ yr$^{-1}$. 
The source IRS6 showed no obvious accretion signature in Nisini et al. (2005) and only an upper 
limit on the accretion rate of $\dot{M_\mathrm{in}}\lesssim5\times10^{-9}$ M$_{\odot}$ yr$^{-1}$ was given.
Among these three sources, the accretion rates and the steady 3.5-cm fluxes seem to be correlated. 
The younger binary source IRS5 is deviated from this correlation (Figure \ref{fig_accretion}), 
which suggests that comparison between different types of sources may not be straightforward. 
We note that the large polarization percentage of IRS5 as compared with IRS7W and IRS7E (Table \ref{tab_stokesV}), 
and the large variation of the spectral index (Figure \ref{fig_index}), 
indicate that a good fraction of the Stokes \textit{I} flux from IRS5 is non-thermal emission. 
Because of the $>$30-day timescale of the Stokes \textit{V} flare observed in IRS5 
(Figure \ref{fig_stokesV}), we think that for this particular source, the non-thermal emission 
cannot be filtered out by the biweight statistics, and thus will contribute significantly 
to the steady flux level.
By observing where the spectral index of IRS5 converges (Figure \ref{fig_index}), 
we hypothesize that the flux of the more stationary thermal emission may be 
at most 0.4 to 0.6 mJy.

\subsection{The Time-Domain Structure Function}
\label{sub_structure}

We compare our observations with the previous observations reported by Feigelson et al. (1998), Forbrich et 
al. (2006), and Choi et al. (2008, 2009).
We note that there were  4 additional epochs of radio observations on 2005 August 09, 10, 12, and 13, 
reported by Forbrich \& Preibisch (2007).
These observations were executed in the most compact VLA D array configuration and can contain
extended emission that hinders a comparison with observations in more extended array configurations.
They may possibly be included in our statistical analysis in the future, after the effect of the 
extended emission is better modeled. 
We do not include these observations in our current analysis.
Nevertheless, the daily flux variabilities provided by these 2005 observations were also well sampled in our 2012 March observations (Table \ref{tab_obs}). 

\begin{figure}
\begin{tabular}{p{9cm} }
	\\
\end{tabular}

\vspace{-2.4cm}

\hspace{-2cm}
\begin{tabular}{p{9cm} }
\includegraphics[width=11.5cm]{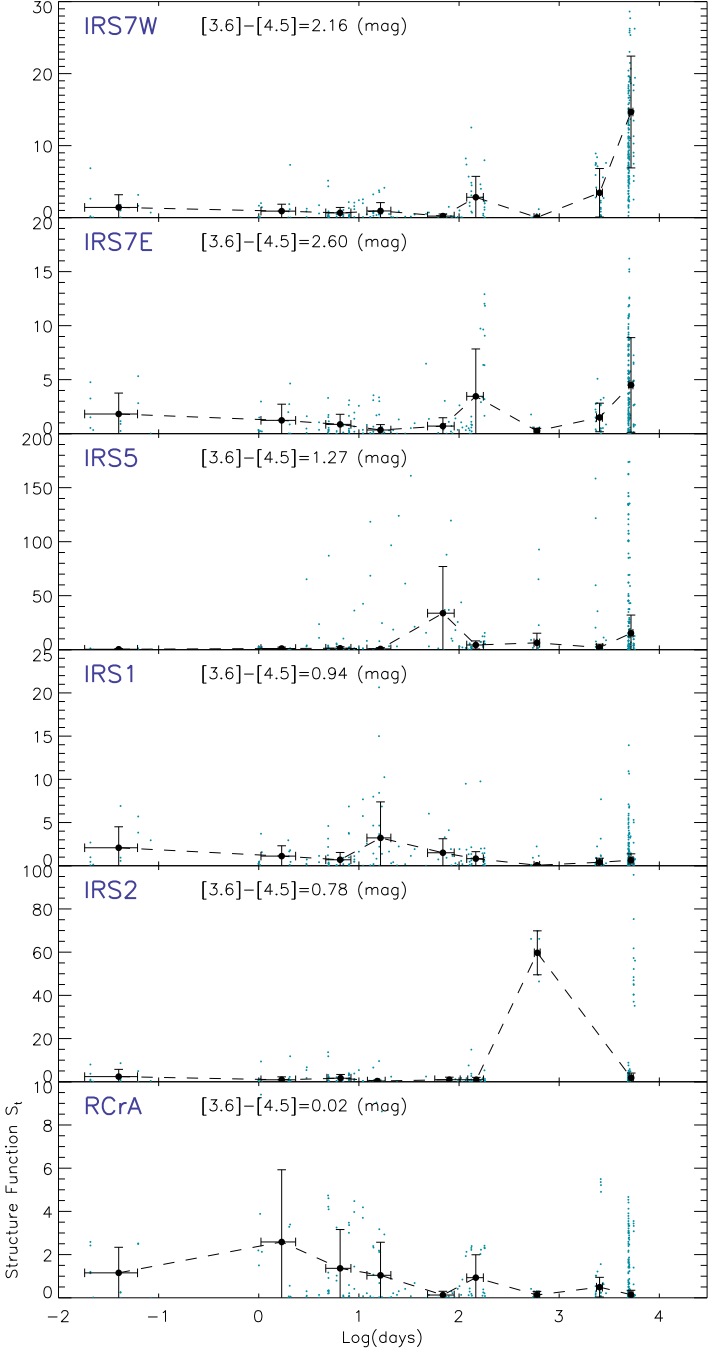} 	\\
\end{tabular}
\vspace{-2.8cm}
\caption{\footnotesize{
The structure function $S(\tau)$ of the frequently or always detectable sources (Section \ref{sub_structure}). 
Green points show the values $S_{ij}$ calculated from the individual pairs of observations. The black symbols show the biweight mean and the biweighted $\pm$1 $\sigma$ standard deviation in the following 9 bins of time-lag (in units of days):  [0.01, 0.1], [1.0, 3.0], [3.0, 10.0], [10.0, 30.0], [30.0, 100.0], [100.0, 300.0], [300.0, 1000.0], [1000.0, 3000.0], [5000.0, 10000.0].
These irregularly separated time-lag bins are chosen to have enough of data points each.
}
}
\label{fig_struc}
\end{figure}

We modified the structure function analysis introduced in Bondi et al. (1994), which was used to derive 
the timescale of variability. 
For each YSO source (Table \ref{tab_sources}), for each pair of data points $i$, $j$, we calculate the 
time-lag $t_{ij}$=$t_{i}$-$t_{j}$, and the normalized flux dispersion 
$S_{ij}$=[$B_{i}$-$B_{j}$]$^{2}$$/$$\sigma_{\mbox{\scriptsize{bw}}}^2$, where $B_{i}$ is the flux observed in 
epoch $i$, and $\sigma_{\mbox{\scriptsize{bw}}}$ is the biweighted standard deviation of the Stokes \textit{I} 
fluxes calculated from 2012 March 15 to September 12.
This analysis can only be performed for IRS7W, IRS7E, IRS5, IRS1, IRS2, and R\,CrA, 
because the other sources are too faint to be detected in individual epochs of the earlier observations. 
The derived $t_{ij}$ and $S_{ij}$ are plotted in Figure \ref{fig_struc}.
Because our sampling of the time-domain baselines is not very uniform, the biweight mean and the biweighted 
standard deviation of $S_{ij}$ can only be calculated in arbitrarily selected bins of time-lag.
The $S_{ij}$ in the time-lag range of [0.1, 1] days cannot be sampled by ground based radio observations. 
For the time-lag bins with relatively poor statistics, possible flaring or fading events can 
dramatically bias the means and the standard deviations of $S_{ij}$.
The most obvious example is observed in IRS2, in which the 0.67 mJy flux on 1997 January 19/20 contributed 
to the large $S_{ij}$ values at $>$300-day timescales.
In fact, the mean and standard deviation of $S_{ij}$ in the time-lag range of [300, 1000] days 
are affected by this poor-statistics issue for all sources. 
In the other time-lag intervals, the behavior of IRS2 is similar to that of IRS1.

We observe that from the top to the bottom panels in Figure \ref{fig_struc} (i.e., from early to late YSOs), 
the timescale of the most significant flux variability shifts from over 1000 days to about 1 day.
In most time-lag bins, the $S_{ij}$ of R\,CrA is consistent with 1 within 2$\sigma$. Its 
long-term flux variability appears to be less significant than 
short-period variations (Section \ref{sub_biweight}). 
Also, we do not find obvious decadal variability in IRS1 and IRS2.
For IRS5, the Stokes \textit{I} flare with a duration of $\sim$30-120 days in 1998 (Choi et al. 2009) leads to the some large values of $S_{ij}$ in the corresponding time-lag bins.
Because of the large circular polarization percentage during the 1998 IRS5 flare event, it is likely to be (at least partially) non-thermal (Choi et al. 2009).
Since no VLA observation was taken between 1998 October 14 and 2005 February 03, we cannot know for how long that IRS5 flare event lasted. 
Figures \ref{fig_flux1} and \ref{fig_struc} consistently suggest that the decadal variability of 
IRS7E is marginally larger than its short-period variabilities, and the decadal variability of IRS7W 
is significantly larger than its short-term variability.
Choi et al. (2008) also suggested that source IRS7W may be undergoing a long duration outflow eruption.

\subsection{Interpretation}
\label{sub_interpretation}
We think that the detected 3.5-cm Stokes \textit{I} emission from the young YSOs IRS7W, IRS7E, IRS5, IRS1, and 
IRS2 is produced by a mixture of thermal radio emission from the jet cores and 
gyrosynchrotron emission from the magnetic reconnection events.  
The measured spectral indices (Figure \ref{fig_index}) provide hints on this. 
Statistical studies on observations of solar flares with durations of 1 to 1000 seconds (e.g., Nita et al. 2004) 
suggested that the centimeter-band spectral energy distribution can be described by the following 
gyrosynchrotron-like spectral shape (Stahli et al. 1990)
\begin{equation}
S(\nu) = e^{A}\nu^{\alpha_{\mbox{\tiny{lf}}}}[1 - \exp(-e^{-B}\nu^{-\beta})],
\label{eq_solarflare}
\end{equation}
where S($\nu$) is the radio flux at a frequency $\nu$, and $A$ and $B$ are parameters affecting the normalization. 
The asymptotic behavior of this spectral function above and below the peak frequency follows 
the positive spectral index $\alpha_{\mbox{\scriptsize{lf}}}$ and the negative spectral index 
($\alpha_{\mbox{\scriptsize{lf}}}-\beta$), respectively (e.g. Lim et al. 1994). 
The detected values of $\alpha_{\mbox{\scriptsize{lf}}}$ and ($\alpha_{\mbox{\scriptsize{lf}}}-\beta$) from the sun can be up to $\sim\pm$6.
At a certain observing frequency, whether one sees positive or negative spectral indices from 
the flares more often depend on the distribution function of the peak frequency. 
Due to averaging during integration times much longer than the duration of the flares, the observed 
spectral indices $\alpha^{\mbox{\scriptsize{obs}}}$ are likely to be closer to those of the most frequent events. 
Regarding the thermal contribution to the flux, 
the spectral index of optically-thin thermal radio emission is $\sim$$-$0.1 (e.g., Anglada et al. 1998).
Optically thicker thermal radio emission can have spectral indices up to $+$2.
From analytical calculations, Reynolds (1986) suggested that the spectral indices of radio jet cores 
range from $+$0.2 to $+$2.

The convergence of $\alpha^{\mbox{\scriptsize{obs}}}_{\mbox{\tiny{IRS7E}}}$, 
$\alpha^{\mbox{\scriptsize{obs}}}_{\mbox{\tiny{IRS7W}}}$, $\alpha^{\mbox{\scriptsize{obs}}}_{\mbox{\tiny{IRS5}}}$, 
and $\alpha^{\mbox{\scriptsize{obs}}}_{\mbox{\tiny{IRS1}}}$ towards 0 or slightly positive values 
during their lower flux status may be consistent with steady emission from the thermal radio jet 
cores (Section \ref{sub_index}, Figure \ref{fig_index}). 
The frequent $\alpha^{\mbox{\scriptsize{obs}}}<$0 gyrosynchrotron flares can be expected if the distribution 
functions of the peak frequency are similar to the case of solar flares, which achieve the maximum 
around 7 GHz (Nita et al. 2004).
This argument is not always valid for IRS5 because part of its non-thermal emission may be from the 
larger scale magnetic field, and thus can have $>$30-day durations.
However, $\alpha^{\mbox{\scriptsize{obs}}}_{\mbox{\tiny{IRS5}}}$ on 2012 March 17 (Figure \ref{fig_index}) varies 
from $\sim$0 to $\sim$2 within 3 hours (see Table \ref{tab_obs}), which may also be attributed to a mechanism 
similar to solar flares.
We do not provide a good constraint on $\alpha^{\mbox{\scriptsize{obs}}}_{\mbox{\tiny{IRS2}}}$, 
besides that it seems to be positive. 
We tentatively think that the emission mechanism of IRS2 is similar to IRS1, because of the similarity 
in the Stokes \textit{I} flux variability (Figures \ref{fig_flux1} and \ref{fig_flux2}, and Section 
\ref{sub_structure}) as well as the similarity in their \textit{Spitzer} colors (Figure \ref{fig_spitzer}).
A non-thermal emission mechanism was also suggested from comparisons between the X-ray and the radio emission associated with YSOs (Feigelson
1998), though the underlying connection is not yet fully understood (e.g., G{\"u}del 2002). 
However, X-ray emission is a clear sign of magnetospheric activity, and in Table 3, only the two radio faint Class 0/I YSOs JVLA2 and JVLA4 were not detected in the earlier deep X-ray survey (Forbrich \& Preibisch 2007). 
These non-detections could, however, be due to foreground extinction.

The Stokes \textit{V} flares detected from IRS7E and IRS7W support the idea that the active magnetospheres 
were developed while some magnetic loops occasionally break through the optically thick radio emission wind 
(Figure \ref{fig_stokesV}).
The very low circular polarization percentages of these two sources as compared with that of IRS5 
(Table \ref{tab_stokesV}) suggest that either the dominant emission mechanism of these two sources is thermal, 
or that the non-thermal emission originated from the obscured inner regions.
This is also supported by the decadal flux variability of these two sources (Section \ref{sub_structure}).
The magnetosphere of IRS5 may be less obscured by a radio jet core, thus its circular polarization 
percentage during the Stokes \textit{V} flares is closer to what usually is observed 
from the gyrosynchrotron sources (Andre 1996). 

The variability in timescales $<30$ minutes of the 3.5-cm Stokes \textit{I} emission from the rest of the sources,  
except JVLA2, JVLA3, JVLA4, and IRS5N (Figure \ref{fig_flux2}), suggests that 
gyrosynchrotron emission is dominant.
Their Stokes \textit{V} emission, unfortunately, could only be detected with S/N$>$2 if the circular 
polarization percentage is much larger than 20\%. 
We are surprised that although the \textit{Spitzer} colors of IRS6 are similar to IRS1 and IRS2 
(Figure \ref{fig_spitzer}), the radio flux and variability of this weakly accreting YSO 
(Section \ref{sub_biweight}) behave more similar to those of the Class II source T\,CrA (Figure \ref{fig_flux1}, 
\ref{fig_flux2}).
The Class III source JVLA1 (CrA\,PMS 1) is also only occasionally detected at 3.5 cm, but shows a peak flux 
$\sim$7 times larger than IRS6 and R\,CrA (Figure \ref{fig_flux1}), and has the highest fractional maximal 
flux deviation among all observed sources (Figure \ref{fig_statistics2}).

Overall, we speculate that the YSOs at the earliest 
evolutionary stage, such as IRS7W and IRS7E, have their large-scale magnetosphere embedded
inside the optically-thick thermal radio jet cores. 
The more extended part of the thermal radio jet can contribute with $\alpha^{\mbox{\scriptsize{obs}}}\sim$0 
emission. 
Bright non-thermal flares can be observed when the large magnetic loops occasionally break through 
the optically thick jet core, or if the jet core is porous. 
The mass loss (and the accretion) of YSOs in the evolutionary stage of IRS5 may be less active than 
for IRS7W and IRS7E.
The non-thermal emission then becomes easier to observe because the jet core is optically thinner 
or more porous. 
At the evolutionary stage of IRS6 and T\,CrA, the YSOs fade in radio emission due to both 
weaker mass loss and a not fully-developed stellar magnetic field.
At later stages, gyrosynchrotron emission becomes dominant after the full development of the stellar 
magnetic field. 
The radio emission of the Class 0/I YSOs in our sample appears to be bimodal: the emission is always
either very bright (e.g., IRS7E, IRS7W, IRS5), or faint (e.g., JVLA2, JVLA3, IRS5N, CrA-24). 
This may imply that the radio emission of Class 0/I YSOs has variability in a much longer timescale than 
what is probed by the radio observations presented in this paper ($\sim$15 years).
Accretion disk instabilities could cause variability on such a timescale. 
A deep infrared spectral-line survey is required to check whether the radio faint Class 0/I YSOs 
are weakly accreting sources.
From inspecting the flux variability of the radio-faint Class 0/I YSOs, it seems that the large-scale 
magnetosphere is more active in the radio when strong mass-loss occurs 
(see also Figures \ref{fig_statistics1} and \ref{fig_statistics2}).
Another possible exception is the luminous radio emission from some FU Orionis stars 
(Rodriguez et al. 1990; Anglada et al. 1994; Vel{\'a}zquez \& Rodr{\'{\i}}guez 2001), which might be 
classical T Tauri stars (Class II) during their quiescent phase (Hartmann \& Kenyon 1996).
Because of the small number of observed YSOs in our study, in particular in the Class II and Class III stages, 
the proposed scenario needs to be verified by more extensive surveys.


\section{Summary and Outlook}
\label{chap_summary}
We performed 8-10 GHz monitoring observations towards the young stellar cluster R\,CrA in 2012, using the JVLA.
Efforts have been taken to ensure that the changes of the JVLA array configurations do not interrupt or hinder 
the analysis of long-term radio flux variations.
We found that for this particularly nearby field, after implementing a cut in $uv$ distance  
$>$4.4 $k\lambda$, the effects of changing the JVLA array configuration are negligible compared with the daily 
and hourly radio flux variabilities.

From comparison with previous observations, radio flux variability was detected in timescales from 
$<$30 minutes, up to $\sim$15 years. 
Our current consensus is that the 3.5-cm radio emission from YSOs is dominated by the active magnetosphere and 
by the thermal emission wind.
The active magnetosphere, which produces hourly and daily radio flux variability, is developed as early as 
when the transition from the Class 0 to the Class I phase occurs.
The thermal wind seems to be correlated with the accretion rate and varies in a longer, 
dynamical timescale. 
The optically thick wind can partially obscure the active magnetosphere during the earliest 
stages, thus alleviating the non-thermal confusion in the diagnosis of the thermal radio jet variability.
In stages later than Class II, the mass loss becomes weak, and the radio flux 
is dominated by gyrosynchrotron emission from the stellar magnetic field.

Our scheme needs to be verified because of the small number of observed YSOs. 
In particular, there is only one detectable Class II YSO and two detectable Class III YSOs in our sample. 
Besides, the radio flux variability of exceptional cases like the FU Orionis objects cannot yet be 
incorporated.
Those exceptional cases might be very important for understanding protostellar evolution.
A more extensive JVLA survey may shed light on these issues.

We also note that the right radio emission mechanisms should show not only the correct timescales, 
spectral indices, or polarization percentages, but also the characteristic flux scales. 
In the future, more sensitive observations using the Square Kilometer Array (SKA) will be important for 
characterizing the physical mechanisms of fainter, shorter duration emission.

\acknowledgments
HBL thanks Dr. Joseph L. Hora for his help and the useful comments while organizing this project. 
R.G.-M. acknowledges funding from the European Community's Seventh Framework Programme (/FP7/2007-2013/) 
under grant agreement No. 229517R. 

{\it Facilities:} \facility{JVLA}


\end{document}